\pdfoutput=1
\documentclass{aastex63}
\usepackage{graphicx}

\usepackage{amsmath, amssymb}
\usepackage{soul}
\received{November 26, 2021}
\revised{February 11, 2022}
\accepted{February 17, 2022}
\submitjournal{ApJ}

\shorttitle{Surface dynamics of intergranular photospheric vortex tubes}
\shortauthors{Aljohani et al.}
\graphicspath{{./}{figure/}}

\begin{document}

\title{New approach for analysing dynamical processes on the surface of photospheric vortex tubes}

\correspondingauthor{Yasir Aljohani}
\email{yaljohani2@sheffield.ac.uk, ymjohani@uqu.edu.sa}

\author{Yasir Aljohani}
\affiliation{Plasma Dynamics Group, School of Mathematics and Statistics, University of Sheffield, Hicks Building, Hounsfield Road, Sheffield, S3 7RH, UK}
\affiliation{Department of Mathematics, Jamoum University College, Umm Al-Qura University, Jamoum, 25375 Makkah, Saudi Arabia}

\author{Viktor Fedun}
\affiliation{Plasma Dynamics Group, Department of Automatic Control and Systems Engineering, The University of Sheffield, Mappin Street, Sheffield, S1 3JD, UK}

\author{Istvan Ballai}
\affiliation{Plasma Dynamics Group, School of Mathematics and Statistics, University of Sheffield, Hicks Building, Hounsfield Road, Sheffield, S3 7RH, UK}

\author{Suzana S. A. Silva}
\affiliation{Plasma Dynamics Group, Department of Automatic Control and Systems Engineering, The University of Sheffield, Mappin Street, Sheffield, S1 3JD, UK}

\author{Sergiy Shelyag}
\affiliation{School of Information Technology, Deakin University, Geelong, Australia}

\author{Gary Verth}
\affiliation{Plasma Dynamics Group, School of Mathematics and Statistics, University of Sheffield, Hicks Building, Hounsfield Road, Sheffield, S3 7RH, UK}

\begin{abstract}
The majority of studies on multi-scale vortex motions employ a two-dimensional geometry by using a variety of observational and numerical data. This approach limits the understanding the nature of physical processes responsible for vortex dynamics. Here we develop a new methodology to extract essential information from the boundary surface of vortex tubes. 3D high-resolution magnetoconvection MURaM numerical data has been used to analyse photospheric intergranular velocity vortices. The Lagrangian Averaged Vorticity Deviation (LAVD) technique was applied to define the centers of vortex structures and their boundary surfaces based on the advection of fluid elements. These surfaces were mapped onto a constructed envelope grid that allows the study of the key plasma parameters as functions of space and time. Quantities that help in understanding the dynamics of the plasma, e.g. Lorentz force, pressure force, plasma-$\beta$ were also determined. Our results suggest that, while density and pressure have a rather global behaviour, the other physical quantities undergo local changes, with their magnitude and orientation changing in space and time. At the surface, the mixing in the horizontal direction is not efficient, leading to appearance of localized regions with higher/colder temperatures. In addition, the analysis of the MHD Poynting flux confirms that the majority of the energy is directed in the horizontal direction. Our findings also indicate that the pressure and magnetic forces that drive the dynamics of the plasma on vortex surfaces are unbalanced and therefore the vortices do not rotate as a rigid body.
\end{abstract}

\section{Introduction}
A number of modern space- and ground-based observational facilities, e.g. the Solar Dynamics Observatory (SDO), Hinode, Solar Orbiter, Swedish Solar Telescope (SST), Dunn Solar Telescope (DST) and Daniel K. Inouye Solar Telescope (DKIST) allow the determination of key information about various plasma flow and wave processes at different time and spatial scales in the solar atmosphere. This solar region is permeated by the magnetic field generated in the solar interior, and advected to the surface by convective motions. The magnetic field is not distributed uniformly, instead it accumulates in various structures differentiated by their transverse size, lifetime, location, etc.  The photospheric plasma layer presents a rich spectrum of dynamics and classes of flows on all spatial and temporal scales. In particular, a key feature of photospheric plasma flows are the vortex motions. Using  high-resolution series of granulation images taken with the SST, \cite{Brandt88} evidenced vortex structures which visibly dominated the motion of the granules in their neighbourhood. They found that the average lifetime of such structures is about 90 minutes. It was also suggested that such vortices, being  common feature of the solar convective zone, can provide an important mechanism for the heating of stellar chromospheres and coronae by twisting the footprints of magnetic flux tubes. Later,  \cite{Bonet_a1}, based on the motion of bright points determined that the vortices observed in the solar photosphere appear in regions of cooled plasma downflows and they can trace well the supergranulation and the mesogranulation. They also found that the surface density of vortices on the solar disk is approximately $0.9\times 10^{-2}$ vortexes per Mm$^2$. 

Vortices in the solar atmosphere have been observed in a wide range of temporal and spatial scale, from granular \citep[e.g. 0.1-1 Mm in diameter,][]{Giagkiozis2018_a1} to meso- and supergranular scales \citep[e.g. 5-10 Mm in diameter,][]{Bonet_a2, Requerey_a1, Chian_2019_a1}. Depending on the analysed scale, vortices will display different lifetimes. For granular scales, \cite{Giagkiozis2018_a1} applied Fourier local correlation tracking (FLCT) \cite{Fisher2008} to intensity maps and obtained a lifetime around 16.5 s and maximum duration around 100 s. In contrast, supergranular vortices can last for a couple of hours \citep{Requerey_a1, Chian_2019_a1, Chian_2020_a2}.

In the photosphere, the evolution of magnetic elements that co-exist with rotational motion is strongly correlated with those vortices, which act to stabilize the magnetic flux \citep{Requerey_a1}.
Recently, \citet{Shetye_a1} suggested that the chromospheric swirl is a flux tube that extends above a magnetic concentration region in the photosphere. This idea is in accordance with the scenario proposed by \cite{WedemeyerBohm_a1}, where the chromospheric swirls and the photospheric vortices are part of the same solar vortex tube.
In the chromosphere, the swirls have a lifetime around 200-300 s \citep{Tziotziou_a1} and are dominated by transverse and rotational motions \citep{Tziotziou_a2}. 

Solar vortex tubes can be spontaneously generated by turbulent convection. In simulations of quiet Sun regions vortices are found along intergranular lanes \citep{Shelyag_2011b, Kitiashvili2012, Moll_a1, Silva_2020}. These structures have an average lifetime of around 80 s \cite{Silva_2021} and a radius between 40 and 80 km \cite{Shelyag_2013b, Silva_2020}. Solar kinetic vortex tubes \citep[][]{Silva_2021} act as a sink for magnetic field, creating magnetic flux tubes that expand with height \citep{Kitiashvili2012, Moll_a1, Silva_2020}. The concentration of magnetic flux leads to a high magnetic field tension, which can prevent the magnetic field lines to be twisted by the rotational motion \citep{Shelyag_2011a, Moll_a1, Shelyag_2013, Silva_2021}. In some cases, twisted magnetic flux tubes appear close enough to flow vortices, leading to magnetic and kinetic vortex structures closely co-existing in regions with high plasma-$\beta$ \citep{Wedemeyer_2014, Rappazzo_2019, Silva_2021}. The vortical motions can still trigger perturbations along magnetic lines that could lead to wave excitation, e.g. \citet{Battaglia_2021}. The vorticity evolution in the magnetised solar atmosphere is mainly ruled by the magnetic field, which also influences the general shape of vortices \citep{Shelyag_2011b}. Based on the analysis of swirling strength, the part of the vorticity only linked to swirling motion \citep{Shelyag_2011a, Cuissa_2020} showed that the magnetic terms in the swirling equation evolution tend to cancel the hydrodynamic terms close to the solar surface, whereas the magnetic terms dominate alone the production of swirling motion in the chromosphere. The magnetic field also tends to have an important role in the plasma dynamics along the whole vortex tube, as the Lorentz force has a magnitude comparable to the pressure gradient \citep{Silva_2020, Kitiashvili_2013}. High-speed flow jets have also been linked to simulated vortex tubes, driven by high-pressure gradients close to the photosphere and by Lorentz force in the weakly magnetised upper solar photosphere \citep{Kitiashvili_2013}.
In general, the averaged radial profile of magnetic field, angular velocity, pressure gradient inside of the vortex tube at the lower chromosphere and photosphere levels show similar behaviour \citep{Silva_2020}. 

There are several methods that can be used for vortex identification and analysis in fluid flows, e.g. \citet{Gunther_2018}. However, in solar physics most of the previous analysis of vortices in the solar atmosphere were based on visual inspection. Automated methods were first applied in the analysis of solar vortices by \cite{Moll_a1}. Authors applied the vorticity strength method \citep{Zhou_a1}, to identify the area dominated by vortex plasma flows in simulated quiet Sun and solar plage regions. Another vortex identification technique is the $\Gamma$-method, which is able to define both the vortex centre and boundary, and has been used to identify vortices in  observational solar data \citep{Giagkiozis2018_a1}. However, one of the stumbling blocks of the $\Gamma$-method is that it carries out the identification based on the topology streamline of velocity fields, which are not an objective quantity \citep{Haller_a1}. In other words, the $\Gamma$-method is not invariant under time-dependent rotations and translations of the reference frame. This may lead to false vortex detection \citep[see e.g.][]{Silva_2018} as well high dependence on corrections made to remove satellite motion from observational data \citep{Gunther_2018}. Further discussions on the importance of objectivity for vortex identification and description of flow topology can be found in \citep{Haller_a1}. 

\begin{figure*}[htp]
 \centering
     \includegraphics[width=\textwidth]{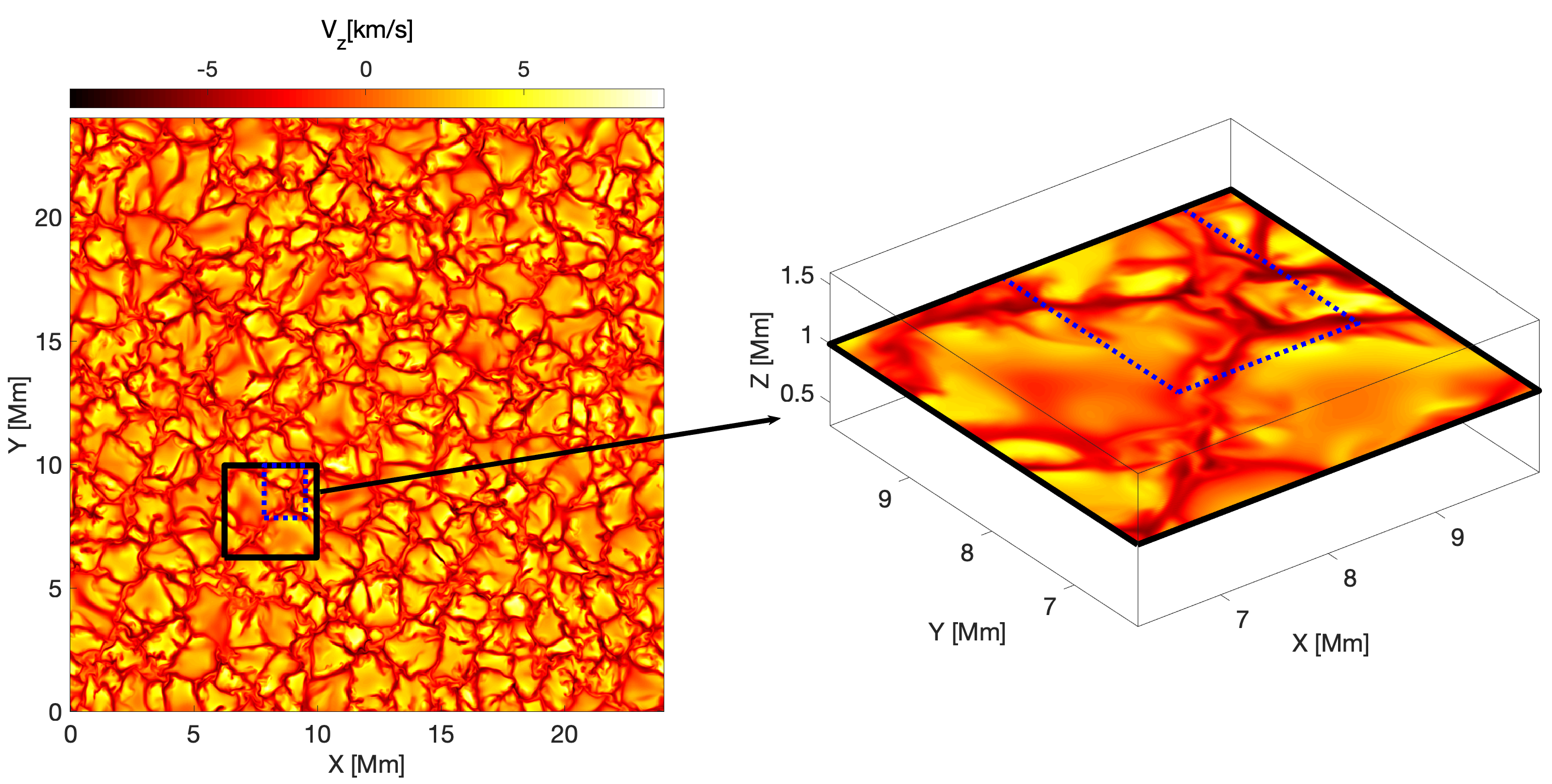}
 \caption{The analysed region of the simulation. Left panel: the $xy$-plane of the whole simulation domain at $z=1.0$~ Mm colored by the $z$-component of the velocity. The selected part of the domain investigated in this paper is delimited by the black square. The blue-dotted square delimits the region used to plot 3D and 2D images of the domain. Right panel: 3D view of the selected part within the black square shown in the left panel.}
\label{domain}
\end{figure*}
The analysis of solar vortices in simulated solar atmosphere data tends to be bi-dimensional due to the limitations of the applied techniques. \cite{Silva_2020} introduced a new methodology to define a 3D vortex based on the Instantaneous Vorticity Deviation (IVD), see e.g. \cite{Haller_a1}, allowing automated detection of boundary and centre of vortex tubes. 
In this paper, we use the method of Lagrangian Averaged Vorticity Deviation (LAVD) developed by \citet{Haller_a1} to identify vortex flows, namely the centre of circulation and their boundary. While the IVD implies an instantaneous field, the LAVD is calculated by advecting and following the particles.  Although they are both based on vorticity deviation, they are distinct methodologies that provide the vortex boundary. The choice for the use of instantaneous or Lagrangian approaches depends on the goal of the study. Here we focus on the understanding of the evolution of plasma dynamics at the vortex boundary and, therefore, LAVD field is more appropriate for the analysis, as IVD only provides the information about a given time frame and does not take into account the motion of the particles. \cite{Haller_a1} compared these two methods and they found that IVD tends to under- or overestimate the size of the vortex at the initial time of the LAVD calculation. IVD also tends to detect short-lived structures, which are not as interesting as long lived ones and fails to determine part of the regions that belong to the true vortex during the time of the analysis. 

The LAVD method is applied in conjunction with MURaM magneto-convection simulation data to detect and track the evolution of 3D vortex tubes in the solar photosphere. The paper is organized as follow: in Section 2 we describe the numerical data, introduce the LAVD technique, explain the isosurface methodology to obtain the vortex tube and present the procedure to project the irregular vortex surface onto the envelope grid. Our analysis of the evolution of plasma variables at the vortex surface is presented in the Section 3. Finally our results are discussed and conclusions are drawn in Section 4.

\section{Methodology}

\subsection{MURaM Simulation Data}

MURaM \citep[MPS - University of Chicago Radiative MHD,][]{2005A&A...429..335V} is a multidimensional MHD code designed to realistically model solar magneto-convection and other related solar photospheric magnetic phenomena, such as pores \citep{2007A&A...474..261C}, sunspots \citep{2009Sci...325..171R}, and flux emergence \citep{2007A&A...467..703C}. 
The code has been used extensively in conjunction with simulated radiative diagnostics to explain a variety of small-scale photospheric phenomena, such as photospheric magnetic bright points \citep{2004A&A...427..335S}, photospheric absorption line profile shapes and asymmetries \citep{2005A&A...442.1059K, 2007A&A...469..731S}, analyse flow structures in photospheric magnetic reconnection \citep{2018A&A...620A.159S} and provide a link between photospheric reconnection events and Ellerman bombs \citep{Shelyag_2013}. MURaM simulations have also been used to analyse the role of torsional motions in the solar atmospheric energy balance \citep{Shelyag_2011a, Shelyag_2012, Shelyag_2013b, Yadav2020, Yadav_2021} and in the analysis of physics and structure of photospheric vortical flows \citep{Shelyag_2011b, Silva_2020,Silva_2021}.

For the purposes of this paper, the code has been set up as follows: the size of the computational domain in Cartesian geometry is set to 24 Mm in the $x$ and $y$ directions and 1.6 Mm in the $z$ direction. The spatial domain is resolved by $960\times960\times160$ grid cells in $x$, $y$, and $z$ directions, respectively. The continuum radiation formation layer (simulated visible ``solar surface'') is located approximately at $z=1~\mathrm{Mm}$ above the bottom boundary, which is made transparent for in- and outflows. Located at the temperature minimum, the upper boundary is closed and allows for the horizontal motion of plasma and magnetic field lines. The lateral boundaries of the simulation box have periodic conditions imposed. The total mass is controlled through correcting the inflow total pressure, deviation of which from the value of pressure at the previous timestep is based on the deviation of the current total mass box from the model, which results in an inflow density change \citep[see][for more details]{2005A&A...429..335V}. On the other hand, the upper boundary allows for horizontal motions and it is located in the higher photosphere, where the density is very low. Therefore, partial reflections from the upper boundary will have only a very small influence on the lower-photospheric layers of the computational domain.

The simulation starts from a well-developed non-magnetic photospheric convection snapshot, where a uniform vertical magnetic field with the strength of $200~\mathrm{G}$ is introduced. The physical reason to choose a uniform vertical magnetic field is because it is divergence and current-free and will therefore not perturb the non-magnetic convection model when it is introduced. Then, the field is advected into intergranular lanes by photospheric flows, and after one granulation lifetime the initial uniformity vanishes. After the magnetic field collapses into the intergranular lanes, the magnetic field concentrations with the strength around $1.5~\mathrm{kG}$ in the photosphere are formed. Then, a series of snapshots, containing state vectors of plasma parameters for each of the grid cells in the domain, are recorded with the cadence of approximately $3.6~\mathrm{s}$. The simulation run contains 120 snapshots, covering roughly 400 seconds of real physical time, corresponding to one granular turnover time.

Figure \ref{domain} displays the $xy$-view of the domain for an $xy$-plane placed at $z=1.0~\mathrm{Mm}$ and colored by the vertical component of the velocity field.
Our analysis is focused on a region located between $x,y = 6.2 ... 9.6~\mathrm{Mm}$, indicated by the black square. This region extends from the simulated surface, $z = 1.0$ Mm, to the lower chromosphere, $z=1.6$ Mm and comprises $150\times150\times60$ grid cells in $x$, $y$, and $z$ directions, respectively. The square marked with blue dotted line is located at {$x,y = 7.8 ... 9.6~\mathrm{Mm}$} and represents the part of the domain used to produce zoom-on view of our results. 

\subsection{Vortex identification}

In this section, we describe the vortex identification in three dimensions based on a sequence of time frames (119 snapshots), for the velocity field provided by the MURaM simulations. Our analysis based on the Lagrangian Averaged Vorticity Deviation (LAVD) method \citep[see e.g.][]{Haller_a1} has been previously applied in solar physics to identify 2D observational vortices \citet{Silva_2018,Chian_2019_a1,Chian_2020_a2}.
As indicated by its name, LAVD is a Lagrangian methodology and it is based on following the particles in order to identify vorticity-dominated regions.  For a plasma velocity ${\bf u(\mathbf{x}(t),t)}$, we can define the vorticity as $\boldsymbol{\omega}=\nabla\times\mathbf{u}$ and, therefore, the LAVD field can be represented as 
\begin{equation}
 LAVD_{t_0}^{t_0+\tau}(\mathbf{x}_0)=
\int_{t_0}^{t_0+\tau}|\mathbf{\boldsymbol{\omega}}
 (\mathbf{x}(t),t)-\langle\boldsymbol{\omega}(t)\rangle | dt.
        \label{eq:2}
        \end{equation}
Here, $\tau$ is a given time interval and $\langle\boldsymbol{\omega}(t)\rangle$ is the spatial mean of the vorticity at time $t$. The position of flow patches, $\mathbf{x}$, is calculated by solving the advection equation,
\begin{equation}
\label{Eq:1}
\centering
    \frac{d \mathbf{x}}{dt}= \mathbf{u}(\mathbf{x},t), 
\end{equation}
over a grid of initial positions $\mathbf{x}_0$ placed in a horizontal $xy$-plane until the final positions $\mathbf{x}(t_0 + \tau)$ are reached within a finite-time interval $\tau$. Note that the LAVD field depends on the integration time. The average lifetime of flow vortices in simulation considered in the present paper is 80 s \citep{Silva_2021}. To perform our analysis we set $\tau=35$ s as after this time the vortex surface becomes deformed and, therefore, further analysis is difficult. This time is roughly half of the average lifetime of a vortex. We used the initial time $t=1321.9~\mathrm{s}$ and integrated until $t=1356.9~\mathrm{s}$.
       
To construct a three-dimensional vortex tube, \citet{Haller_a1} proposed a method based on the isosurface of the LAVD field. First, one identifies the outermost convex contour of LAVD at a chosen height and then finds the isosurfaces, i.e. the set of points having equal values for the LAVD field at all the planes for which LAVD was computed. For our analysis, we have computed LAVD field for the 60 $xy$-planes above the simulated solar surface. The isosurface provides the 3D vortex boundary and describes the initial location for the material elements that undergo the same intrinsic dynamic rotation. The vortex boundary as defined by LAVD, is a rotational Lagrangian coherent structure. Following \citet{doi:10.1146/annurev-fluid-010313-141322}, a Lagrangian vortex can be described as a Lagrangian coherent structure since its boundary is a material surface separating regions with vortical and non-vortical dynamics. In other words, the surface remains together for the time of the analysis, being  defined by the plasma flow and separates the region in the solar atmosphere where the vorticity dominates the plasma dynamics.

We allow some deviation from convexity for the vortex contour as, in a non-ideal fluid, the vortices will most likely have cross-section deviating from circular shapes. This convexity deviation is called convex deficiency and is defined as 
\begin{equation}
c=\frac{A_{c}-A_{ch}}{A_c},
\end{equation}
where $A_c$ is the area which is enclosed by the extracted contour, and $A_{ch}$ is the area enclosed by its convex hull.
Originally, the convexity requirement was established as a way to dismiss false detections caused by high vorticity concentration driven by shear flows. As shear regions do not present convex shapes in non-magnetized flows, the condition of a convex contour would identify only the true vortical motions. However, this requirement is not enough to dismiss wrong vortex identification in the solar atmospheric flows \citep[see e.g.][]{Silva_2018}. Nevertheless, the convex contour also helps to identify the stable vortices as the sturdy tubular vortical structures in the flow present near-circular cross-section. The convexity condition ensures that any material vortex starts out unfilamented at the initial time $t_0$.

\begin{figure}[htp]
 \centering
  \gridline{\fig{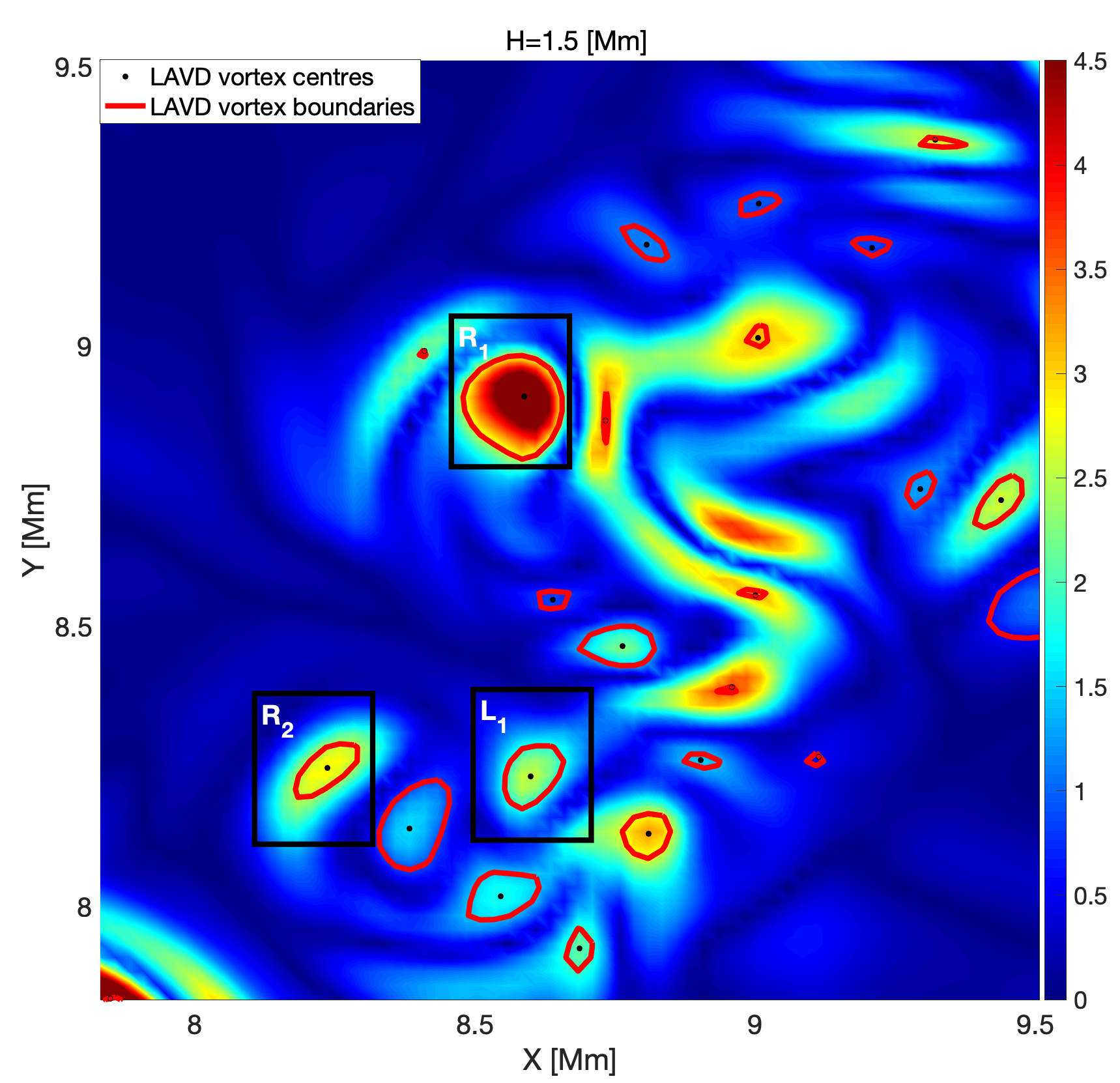} {0.55\textwidth}{(a)}}
   \gridline{\fig{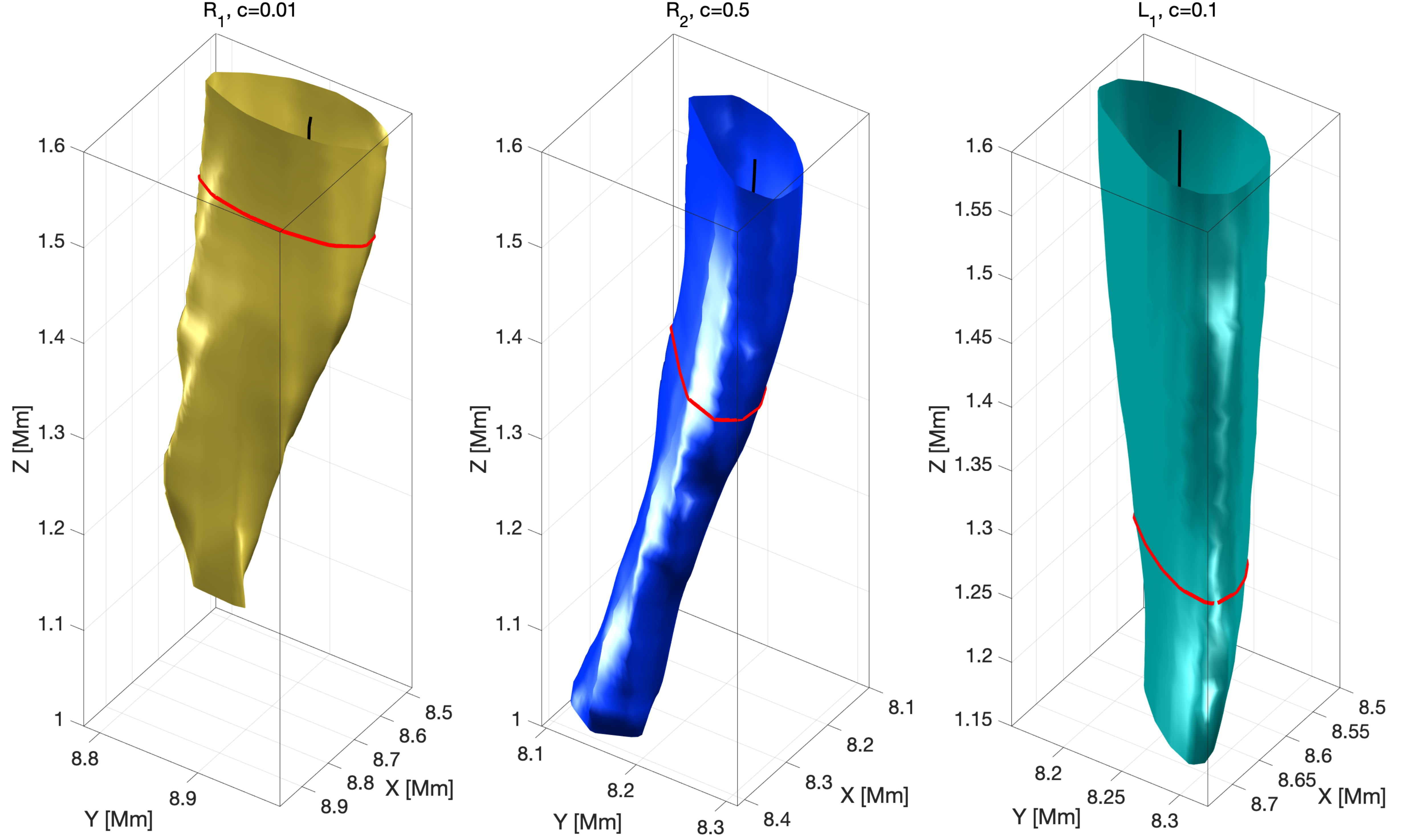}{0.55\textwidth}{(b)}}
 \caption{The top panel shows a two-dimensional horizontal slice of the computational domain at the height $z=1.5$ Mm. The colours correspond to the local values of the LAVD field (see Eq.~\ref{eq:2}). The red contours show the outermost convex boundary of the local maximum of the LAVD field. The selected vortices are located within black squares and labelled as $R_1$, $R_2$ and $L_1$. The bottom panels show their 3D reconstruction (shown in different colors). The cross-section, indicated as a red curve for each 3D surface, represents the selected height to plot the isosurface.}
 \label{fig:LAVDand3vortices}
 \end{figure}

The LAVD field computed at $z=1.5$ Mm is shown in Fig. \ref{fig:LAVDand3vortices} (top panel). The 2D boundary of the identified vortices at this height are depicted by red curves. We select three particular vortices (located within black squares) to analyse further their kinematic and dynamic properties. They are labeled as $R_1$, $R_2$ and $L_1$, where labels $R$ and $L$ represent a clockwise and counter-clockwise directions of their rotation, correspondingly. 
 In general, the whole domain under consideration consists of more than a hundred identified vortices. In order to demonstrate the full potential of LAVD, we focused our analysis only on the vortex tubes that had the greatest vertical length. The selected $R_1$, $R_2$ and $L_1$ vortices spanned over a distance covering both the photosphere and bottom part of chromosphere, their life-times were long enough and they represented the general vortices behaviour features identified in numerical simulation. 
 \begin{figure*}[htp]
 \centering 
 
      \includegraphics[width=0.8\textwidth]{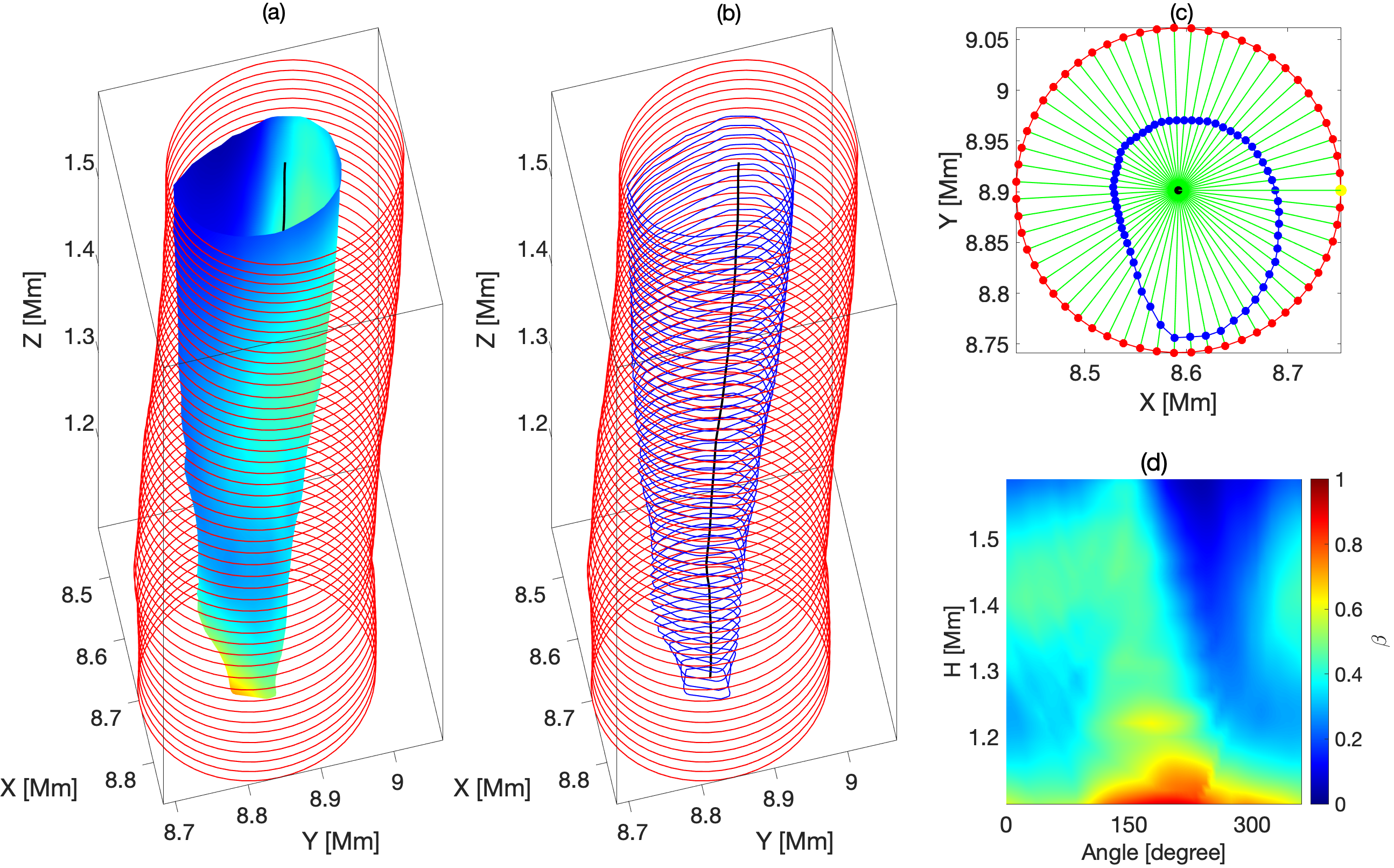} 
\caption{The projection technique used to create a 2D regular surface from 3D irregular shape. (a) The 3D surface displays the vortex tube colored by the plasma-$\beta$. The red circles at each height have centres which come from vortex identification and the position of these are given by the black line. The red circles have the same radius at each height. (b) The blue contours show the boundaries of the vortex. (c) The green segments represent the radius of the vortex and the circle at one particular height. The black dot indicates the centre of the vortex. The vertices of the vortex are indicated by blue dots. The vertices of the circles are indicated by red dots. (d) The last panel describes the projection of the 3D vortex surface in a 2D surface coloured by the plasma-$\beta$. The corresponding video can be found on the  \href{https://sites.google.com/sheffield.ac.uk/pdg/visualisations\#h.73wyu5789fni}{PDG visualisations} web-page.}
\label{reconstruct}
\end{figure*} 
To connect rotational motion from different heights, one should optimize the detection for each vortex, that is the detection should identify the vortex surface such as it encompasses most of the vertical extension of the simulated atmosphere. This can be carried out by trying to define the LAVD contour and distinct values of convex deficiency at different height levels. 
The three-dimensional boundary of $R_1$, $R_2$, and $L_1$ vortices, were established from isosurfaces for the LAVD value at the outermost convex two-dimensional contour at heights of 1.57, 1.44 and 1.32 Mm, respectively.
For each isosurface, a different value of convex deficiency was applied in order to optimize the area encompassed by the boundary. For the chosen vortices the associated values of convex deficiency are $c=0.01$, $0.5$, and $0.1$, respectively. 
The center of each vortex was defined as the local maxima of the LAVD field \cite[see e.g.][]{Haller_a1}. The vortex center is shown as a black dot, see Fig.\ \ref{fig:LAVDand3vortices}(a), or as a black line, see Fig.\ \ref{fig:LAVDand3vortices}(b).

\subsection{Vortex tube projection on a envelope grid}

In this section, we introduce a new methodology to study the plasma dynamics at the surface of the vortex. Due to the complexity of spatial distribution of vortex parameters, e.g. velocity, pressure, temperature, etc. the analysis of three-dimensional vortices is difficult. One possible way is to create the stereoscopic projection of the vortex surface, i.e. mapping of the vortex surface onto a two-dimensional grid (envelope grid), as displayed in Fig.\ \ref{reconstruct}. The three-dimensional view of the original surface of a selected vortex is depicted in Fig.\ \ref{reconstruct}(a). Its surface is colored by the plasma-$\beta$ value and co-centered with the envelope grid represented by the red circles equidistantly distributed in the vertical direction. The values of the variables at the vortex surface were determined by the ``linear'' and ``nearest'' interpolation methods implemented in the  \textit{griddedInterpolant} Matlab function. Both methods were used, but the ``nearest'' approach has an advantage, as this method is faster and requires the least computational memory (see Matlab library for details). Figure  \ref{reconstruct}(b) shows both the boundary and envelop grids. The shape of the vortex surface (blue dots) and its projection onto the envelope grid (red dots) at $z=1.18$ Mm are shown on the panel (c) of the same Fig.\ \ref{reconstruct}. The black dot indicates both, the vortex centre and the centre of the envelope grid (represented by the red circle). The green lines connect the centre of the vortex, corresponding vertices and projected vertices in the envelope. This process is repeated for each height of the vortex to obtain the full projection of the 3D vortex surface on a 2D envelope. The result is shown in Fig.\ \ref{reconstruct}(d), where we display (as an example) the distribution of the plasma-$\beta$. The vertical axis covers the full height of the vortex, while the horizontal axis covers the whole $360^\circ$ range around the perimeter of the lateral surface.

\section{Results}
\begin{figure*}[htp]
 \centering
  \gridline{\fig{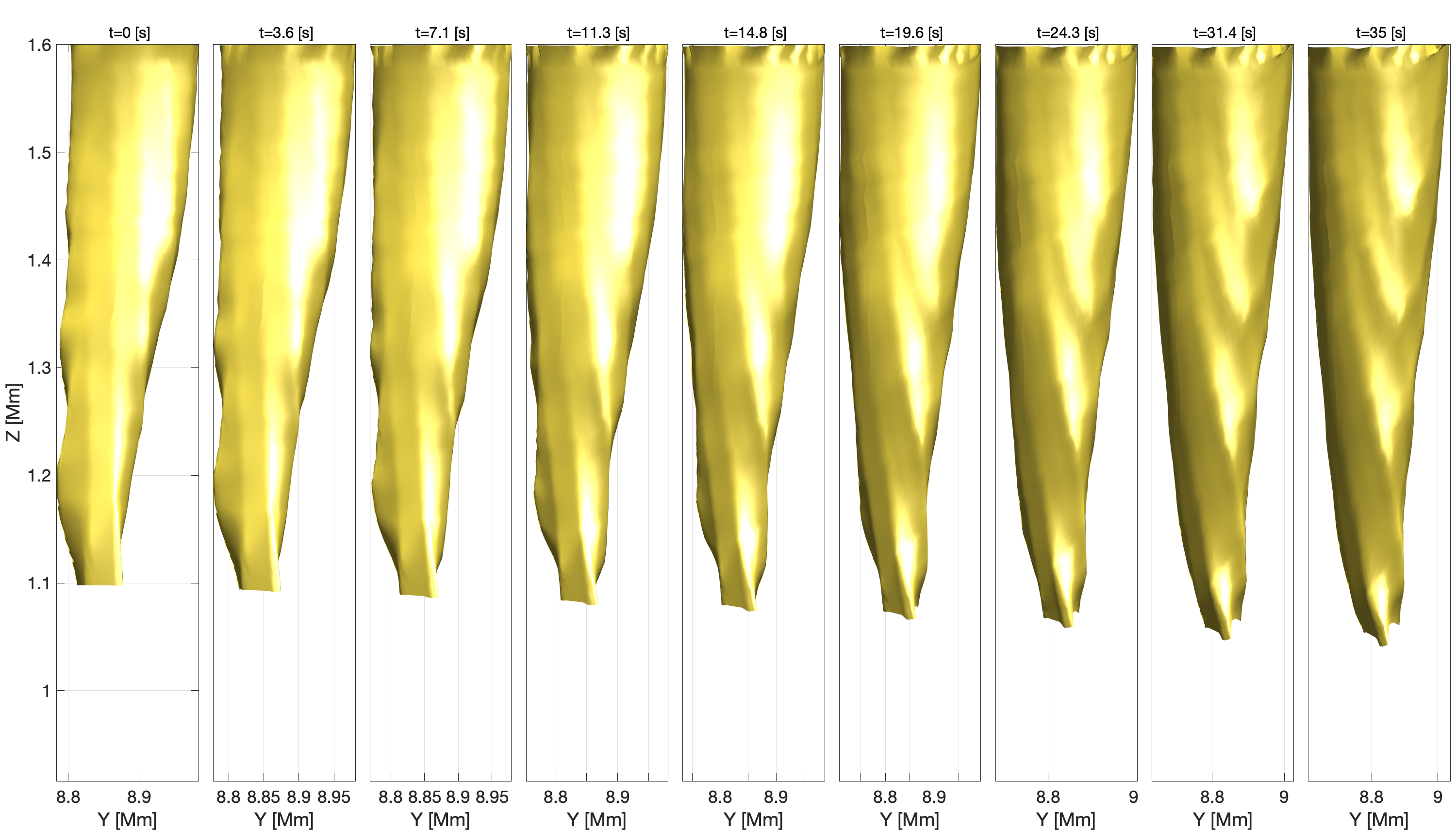}{0.65\textwidth}{(a)}}
    \gridline{\fig{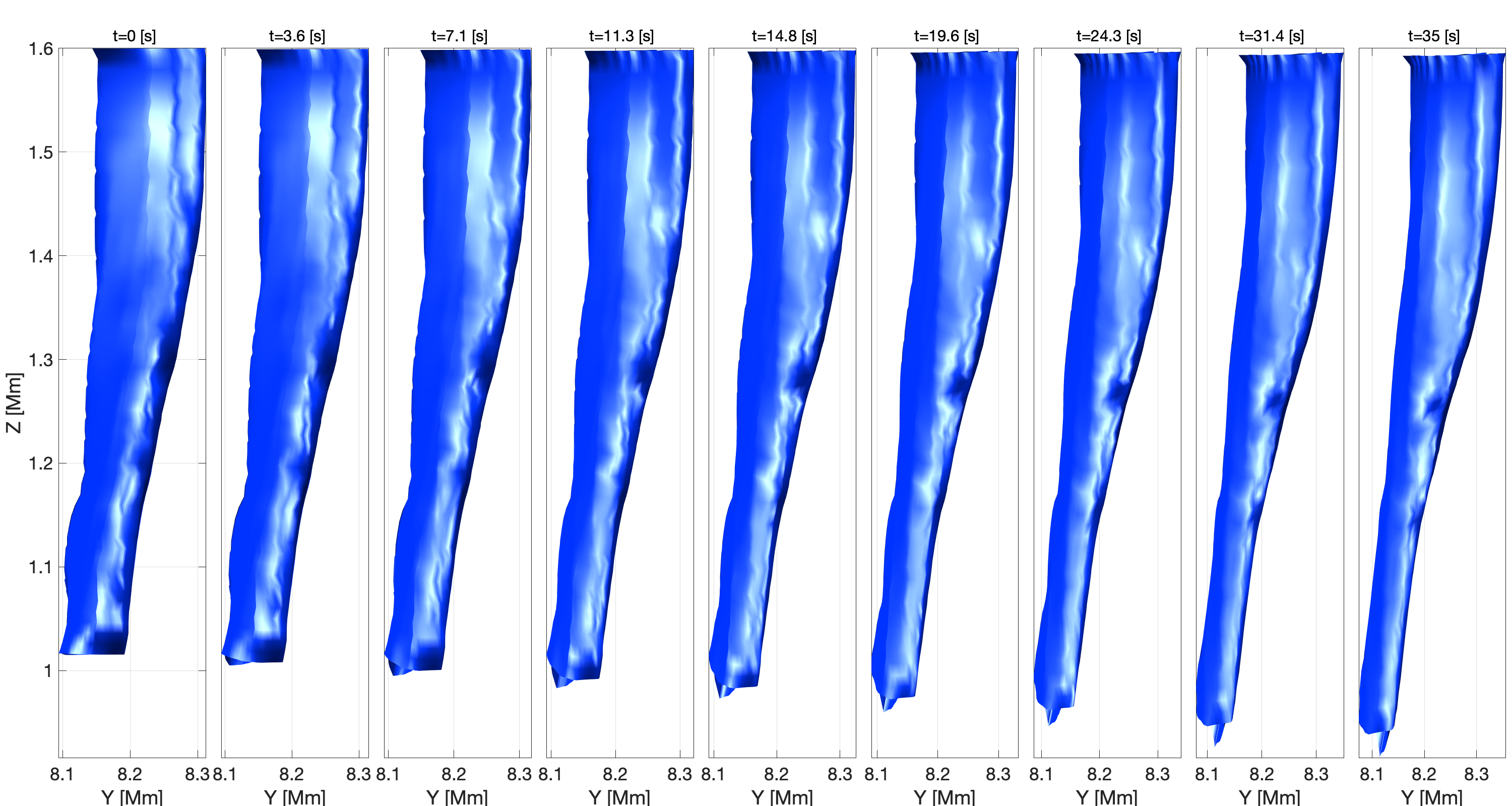}{0.65\textwidth}{(b)}}
  \gridline{\fig{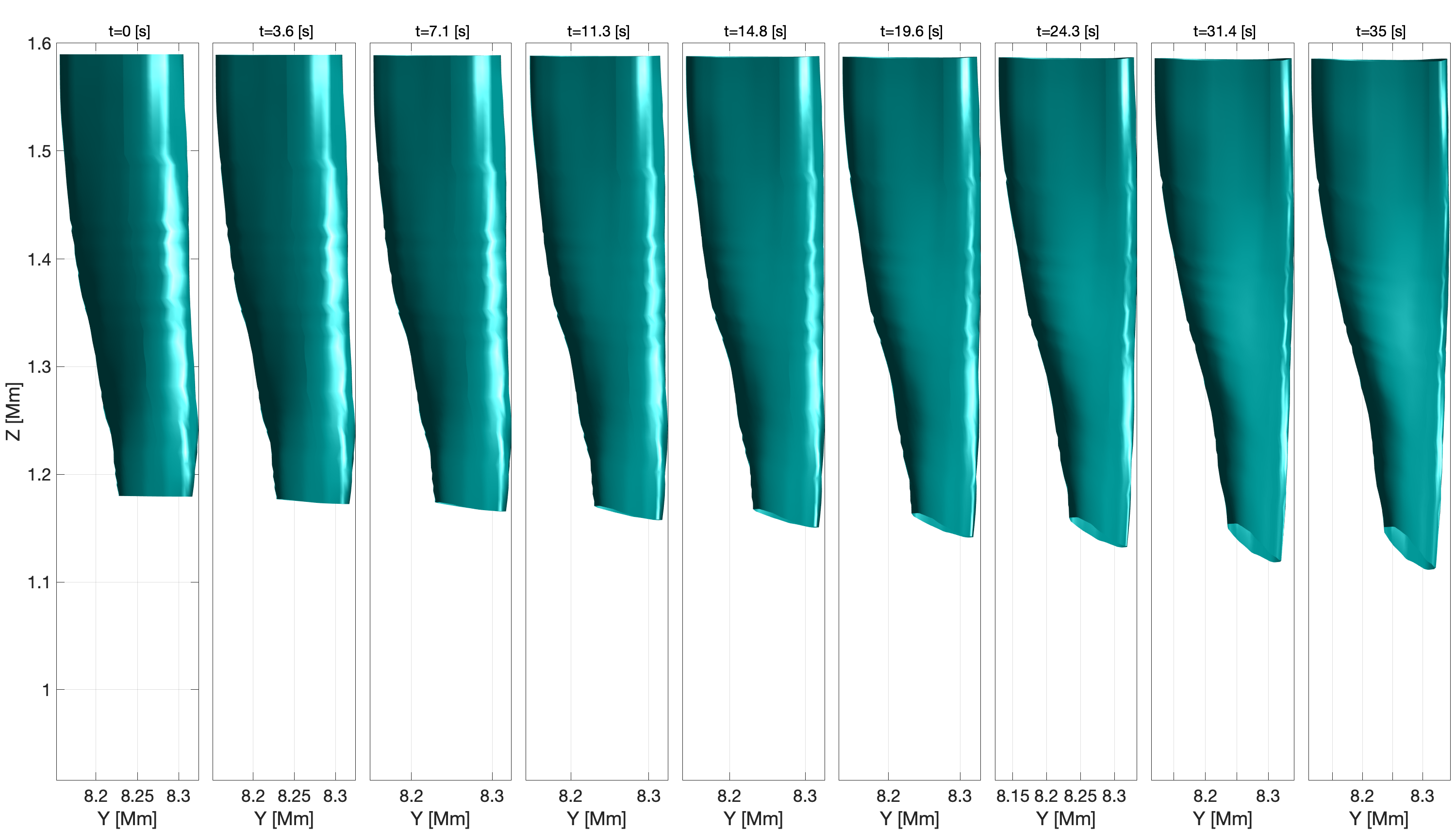}{0.65\textwidth}{(c)}}
\caption{Advected vortex boundaries observed from the $yz$ perspective. From top to bottom: material surface of the $R_1$, $R_2$ and $L_1$ vortices, respectively. The temporal evolution of the three vortices is shown (from left to right) as a time sequence. The different colours are used to identify the vortices.}
\label{advect}
\end{figure*}

The changes on the vortex tubes due to the flow dynamics were established by advecting the particles located at the vortex boundary and saving their position for each time frame. The time interval used for the advection is the same one applied for the LAVD computation, i.e. $t=1321.9~\mathrm{s}$ to $t=1356.9~\mathrm{s}$. The temporal evolution of the vortex's shape for the analysed vortices is displayed in Fig.\ \ref{advect}.
All the structures display the same general tapered tube shape with some deformations along their surface. Figure \ref{advect} indicates that, as the vortices rotate, they tend to be stretched in the vertical direction due to the existing downflows in the intergranular lanes. 
\begin{figure*}[htp]
 \centering
    \includegraphics[width=0.9\textwidth]{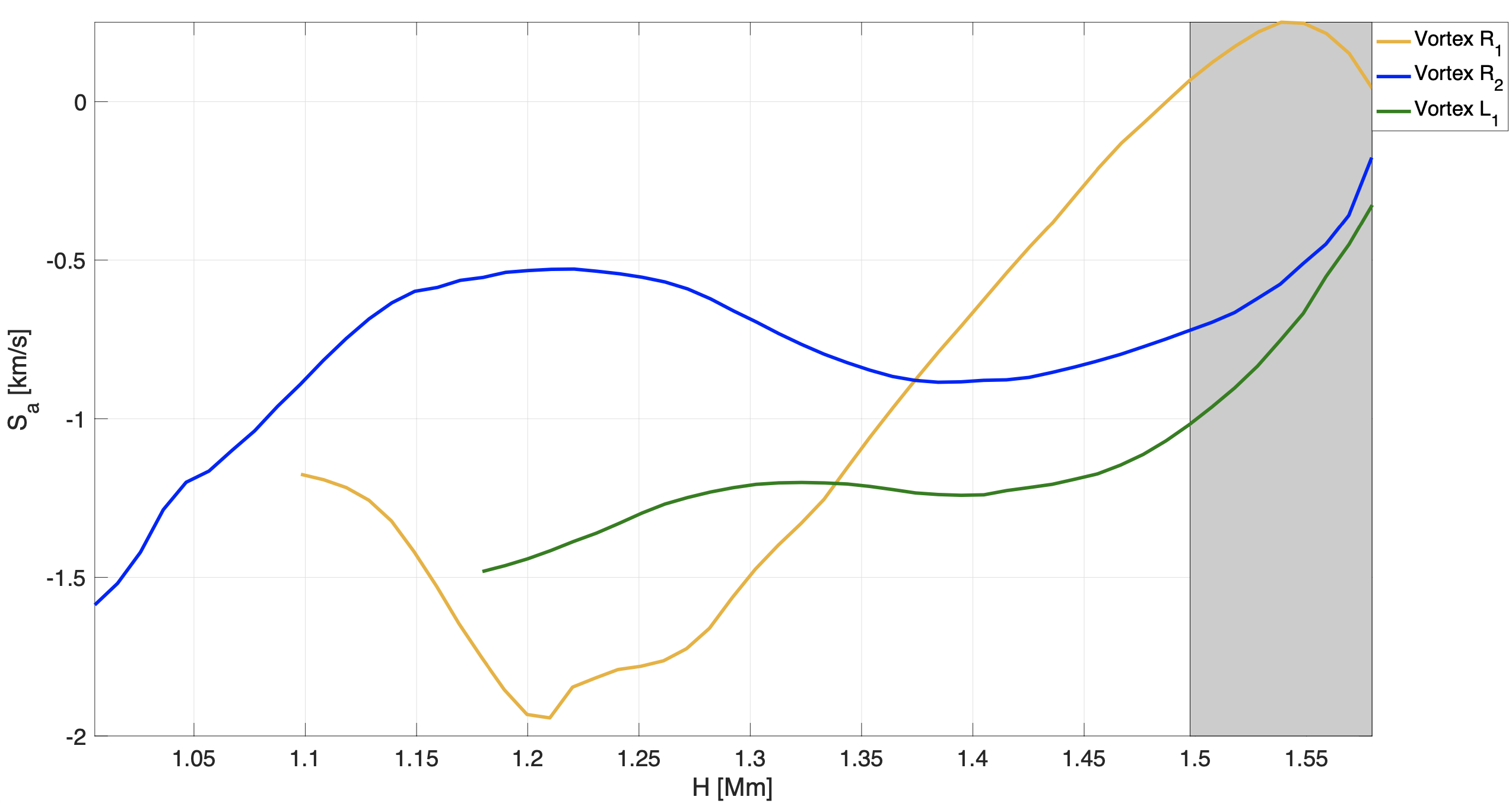} 
\caption{The average downdraft speed ($S_a$) as a function of height. The physical significance of the negative values in average speed demonstrates that the flows are in the direction of $z$. The range of heights that are near to the top boundary of the MURaM simulation is represented by the grey colour. The values corresponding to the three vortices are shown by  distinct colours.}
\label{Fig:speedflow}
\end{figure*}
To estimate downdraft speed, first, the position of every particle in the vertical direction was calculated for every time snapshot and then propagated across all the advected particles. As a result, the estimated value is in the range 0.7-1.08 km s$^{-1}$. The downdraft speed has large values close to the bottom part of the vortex structures and it tends to zero close to the upper region, as a consequence of the boundary conditions of the simulated domain (see Fig. \ref{Fig:speedflow}). It is clear that the vortex surface tends to present greater stretching in the lower part of its boundary.

\begin{figure*}[htp]
 \centering
    \includegraphics[width=0.8\textwidth]{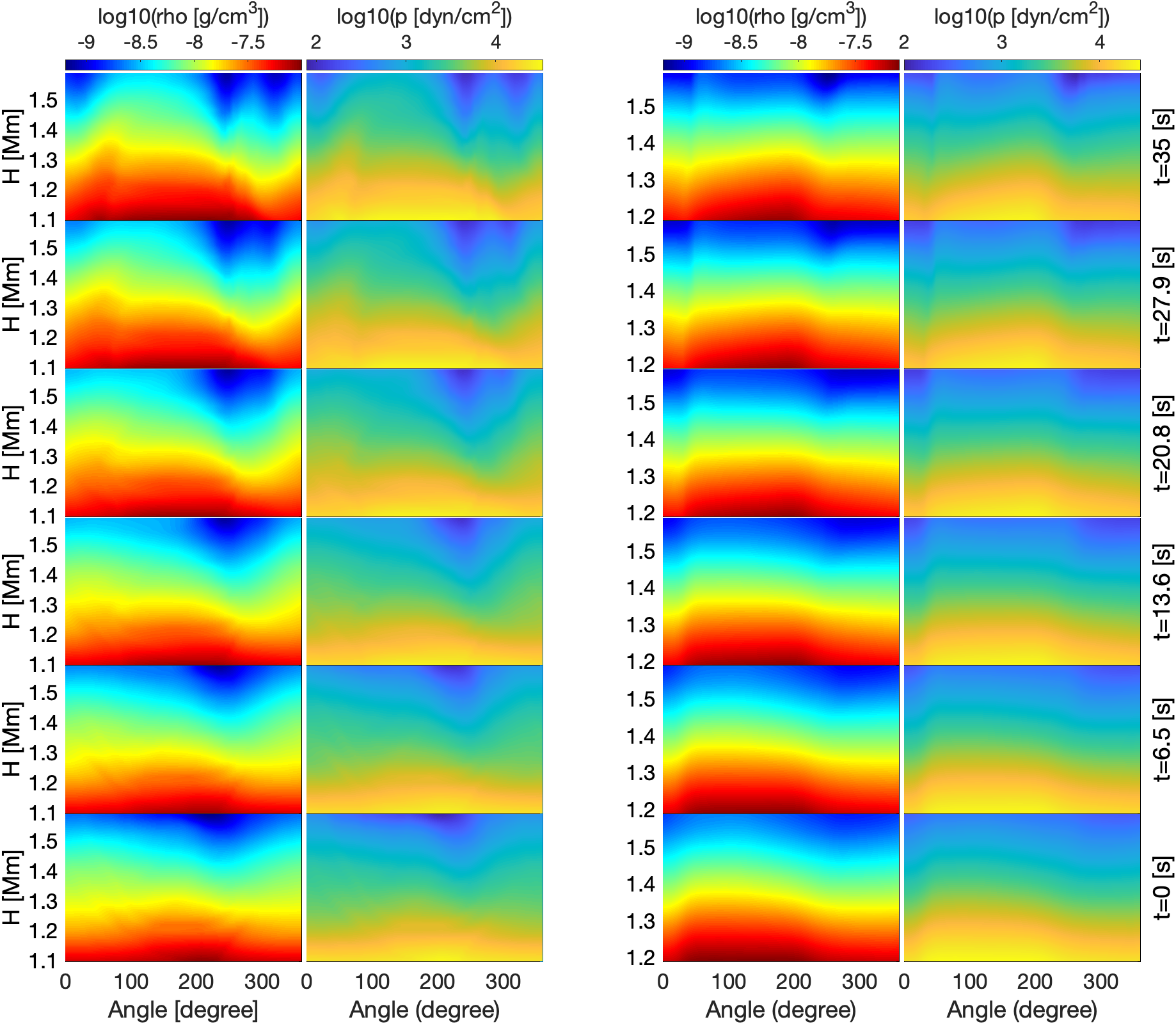} 
\caption{Density ($\rho$) and pressure ($p$) changes with time for vortices $R_1$ and $L_1$, from bottom to top row. The first and second columns represent the $R_1$ vortex, while the final two columns explain the $L_1$ vortex.}
\label{Fig:statevariables}
\end{figure*}

\begin{figure*}[htp]
  \centering
 \includegraphics[width=0.8\textwidth]{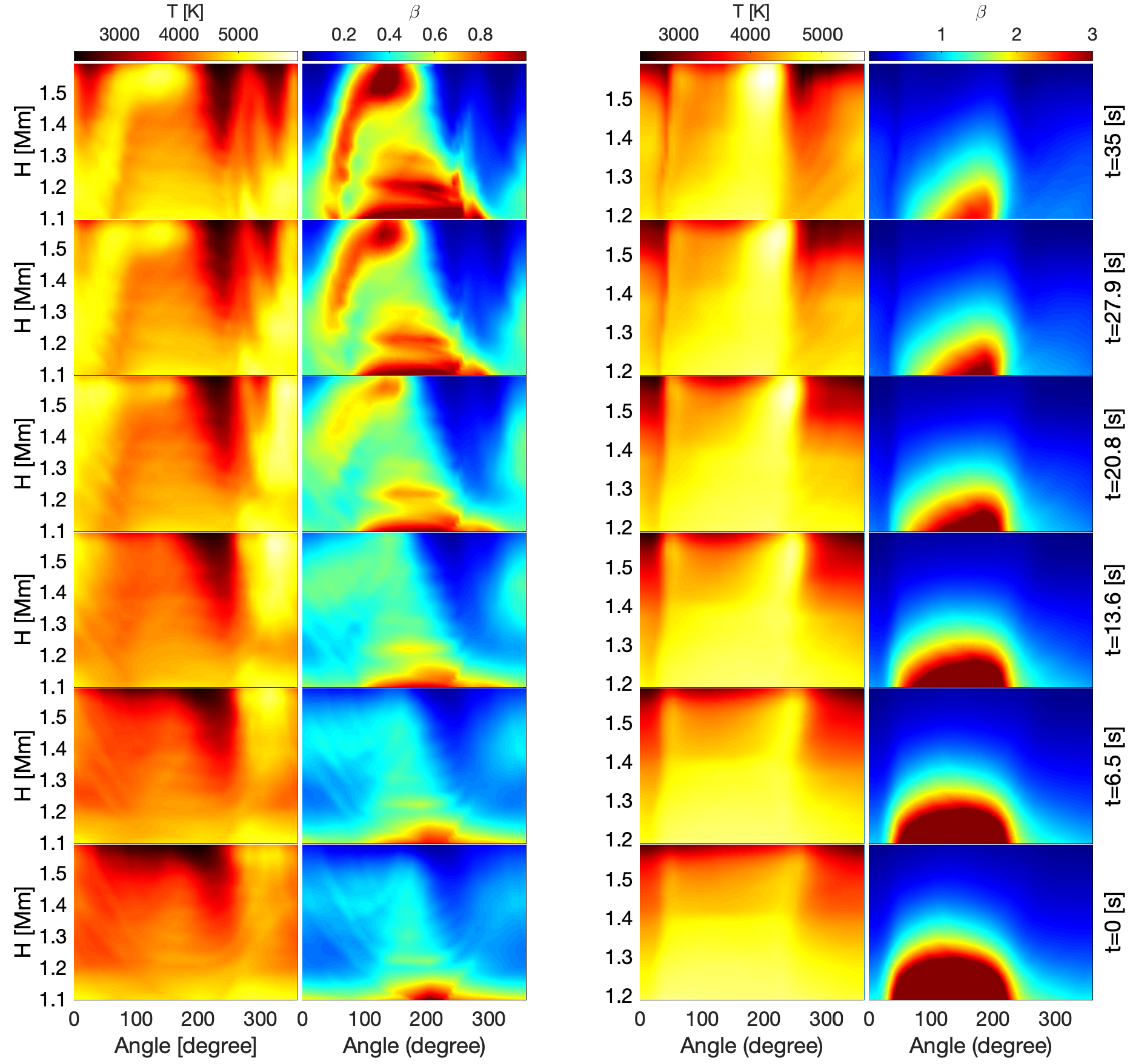} 
\caption{Temporal development (from bottom to top) of temperature ($T$) and plasma-$\beta$ on the surfaces of the $R_1$ and $L_1$ vortices. The first and second columns correspond to the vortex $R_1$, whereas the third and fourth columns correspond to the vortex $L_1$.}
\label{fig:plasma}
\end{figure*}

\begin{figure*}[htp]
 \centering
     \includegraphics[width=0.9\textwidth]{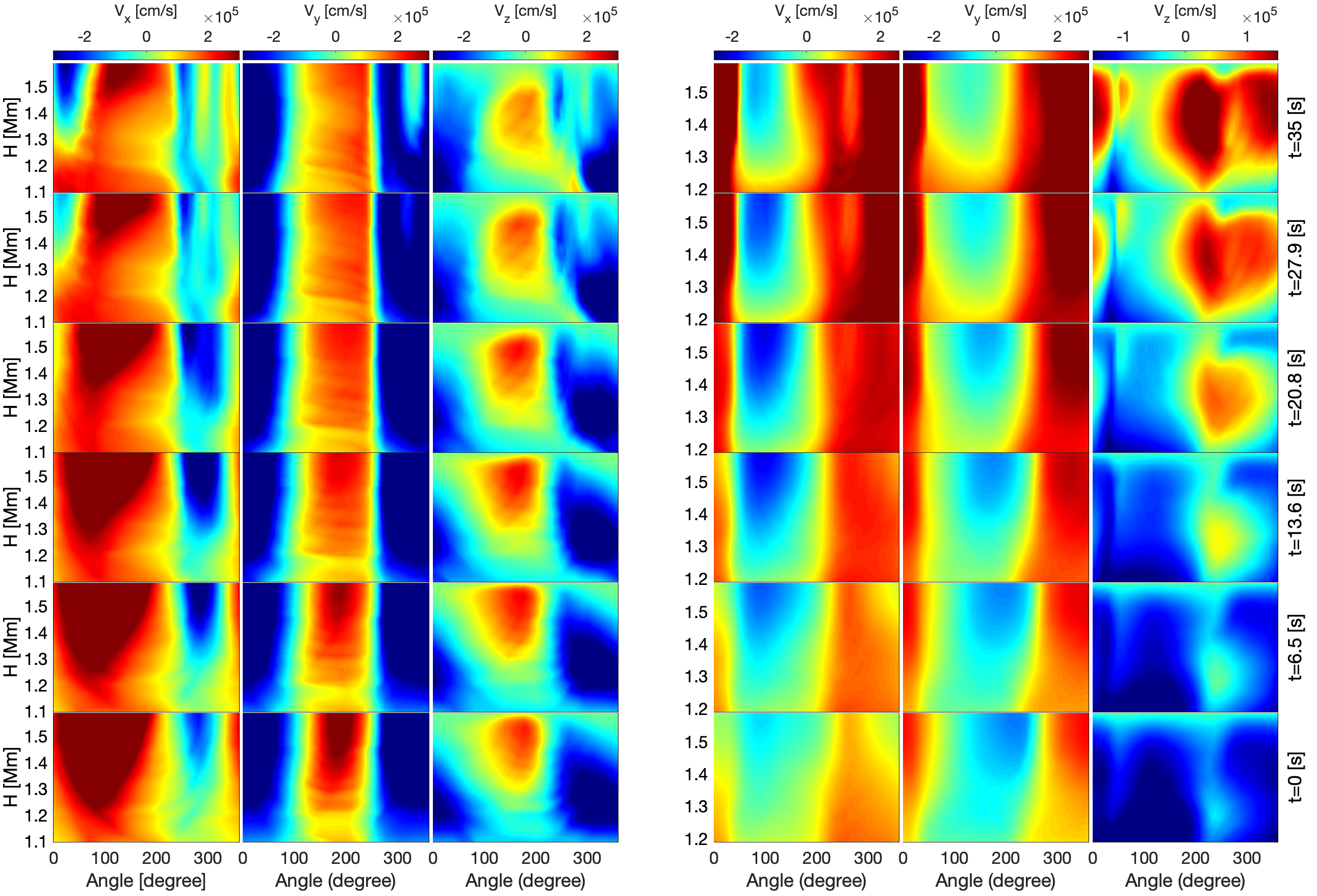} 
\caption{The temporal evolution of the three velocity components ($V_x$, $V_y$, and $V_z$) in the case of vortices $R_1$ and $L_1$. The first, second and third columns stand for the vortex $R_1$ and the last three columns describe the vortex $L_1$.}
\label{fig:velocity}
\end{figure*}

\begin{figure*}[htp]
 \centering
       \includegraphics[width=0.9\textwidth]{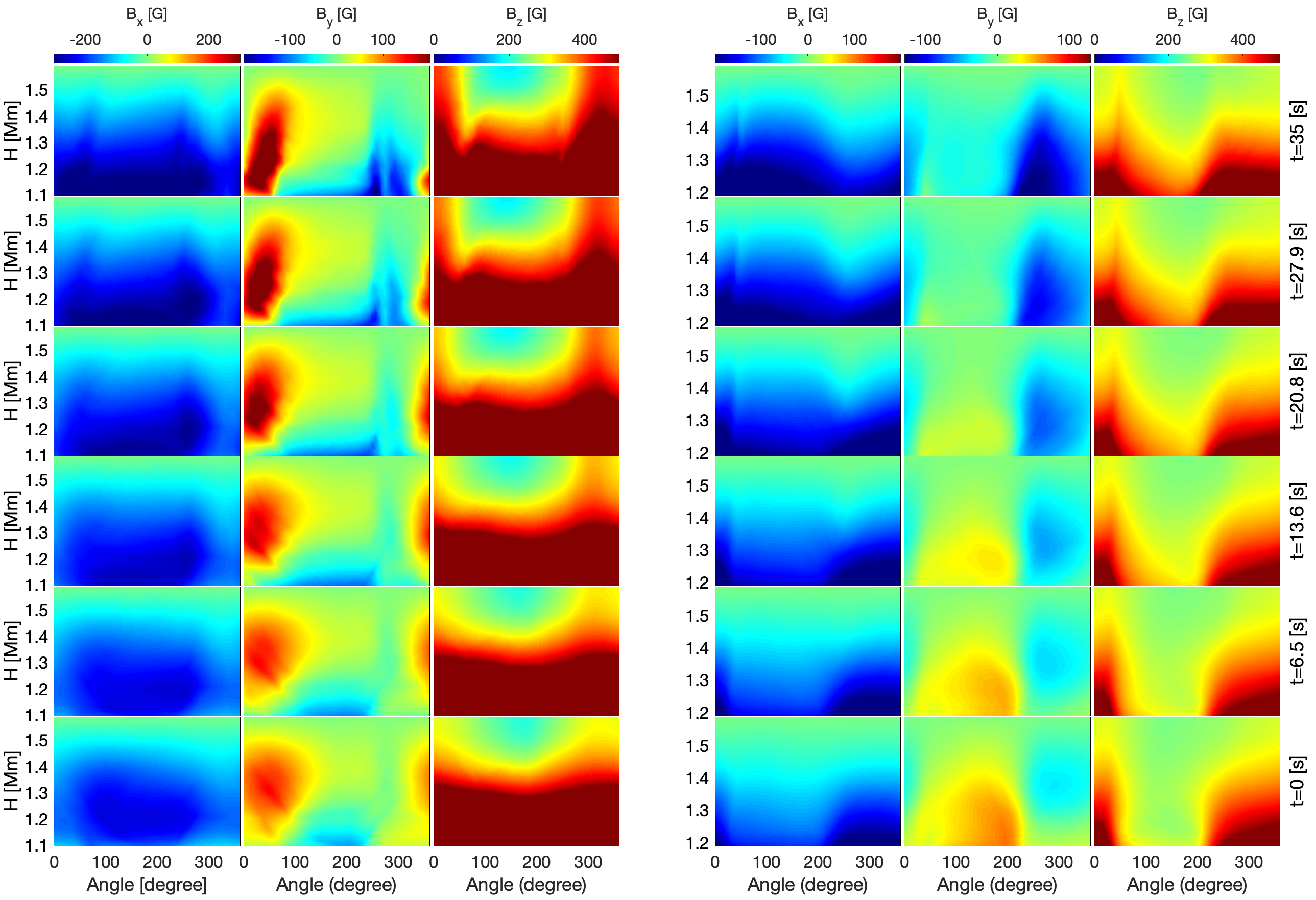}
\caption{The temporal evolution of the three components of the magnetic field ($B_x$, $B_y$, and $B_z$) in the case of vortices $R_1$ (left panel) and $L_1$ (right panel). Since the $z$-component of the magnetic field is always positive, in the colour bar we use only positive values.} 
\label{fig:magnetic}
\end{figure*}

The surfaces of all three vortices can be successfully recovered using the LAVD technique, even in the case of thin structures, such as the vortex $R_2$. However, further analysis in the case vortex $R_2$ is very difficult due to the limited spatial resolution, therefore, in what follows we focus our analysis on the vortices $R_1$ and $L_1$ only. The plasma dynamics during the time interval of the analysis is described by the key plasma variables obtained from interpolation as the particles on the vortex boundary were advected. We carry out a Lagrangian analysis on the vortex evolution, i.e. we study the temporal evolution of the plasma variables at the material surface defined by the particle advection. Our study was performed for the time interval used to compute LAVD, from $t= 1321.9~\mathrm{s}$ to $t= 1356.9~\mathrm{s}$. Figures \ref{Fig:statevariables}, \ref{fig:plasma}, \ref{fig:velocity} and \ref{fig:magnetic}  display the density, pressure, plasma-$\beta$, temperature, velocity and the components of the magnetic field at the surfaces of the $R_1$ and $L_1$ vortices. The panels (from the bottom to the top) show the variation of these physical variables on the vortices' surfaces at six different time-frames, with a regular cadence of 6 s. The vertical extent of each snapshot covers the whole height of the vortex, while the horizontal extent of each snapshot covers the whole $360^\circ$ around the perimeter of the lateral surface. In Figs.\ \ref{Fig:statevariables} and \ref{fig:plasma}, the results obtained for $R_1$ are shown in the first two columns, while the last two columns display the results obtained for the vortex $L_1$.


Figure \ref{Fig:statevariables} displays the Lagrangian evolution of density and pressure for the two vortices. Comparing the evolution with height and time of these two quantities, there is no considerable difference between the two vortices except for small localized variations. Over time, there is a slight change in the density and pressure gradient along the vertical direction. For regions close to the simulated surface, we see the trend to have local concentration of plasma density and pressure as a function of time, but these tend to decrease with height.  

From energetic and dynamical point of view it is essential to study the variation of the temperature and plasma-$\beta$ parameter along the two vortices. The first and third columns of Fig.\ \ref{fig:plasma} show the spatial and temporal evolution of temperature, measured in $K$. One very important result is that the temperature does not have a global behaviour, changes in this important quantity are rather localised. In general the top of the vortices are cooler than their bottom. It is also interesting to note that the surface temperature of vortex $R_1$ is decreasing in those regions where we have a depletion of plasma density. 

The second and fourth columns of Fig.\ \ref{fig:plasma} show the spatial and temporal changes in plasma-$\beta$. One important result visible in these snapshots is that the plasma-$\beta$ is mostly less than one, meaning that the dynamics of the plasma is driven mainly by magnetic forces. It is also clear that during the evolution of the vortex, the value of plasma-$\beta$ increases, in a similar way as the increase in temperature seen in Fig.\ \ref{fig:plasma}. In the case of vortex $L_1$ the dynamics in the bottom part of the vortex is driven mainly by pressure forces, however, in time this diminishes and magnetic forces become more and more dominant. 

The evolution of the three components of the velocity field indicates a stable flow configuration as displayed in Fig.\ \ref{fig:velocity}. This suggests that the detected vortices are stable rotating structures. The variations in the values of the horizontal velocity components suggest the actions of forces acting to accelerate or slow down the rotational plasma motion. As found in previous studies \citep{Kitiashvili_2013,Silva_2020}, the vortices experience both up and downflows during the time of the analysis, as indicated by the evolution of the $z$-component of the velocity.

Figure \ref{fig:magnetic} show the spatial and temporal evolution of the three components of the magnetic field. While the $x$ component of the magnetic field ($B_x$) shows a fairly homogeneous variation towards the bottom of both $R_1$ and $L_1$ vortices, it decreases towards the top. The $B_y$ component shows a strong shear, however, the variation of $B_z$ suggests that the magnetic field decreases with height, creating strong vertical gradients. It is also clear that the vortices have predominantly vertical magnetic component, oriented upwards which increases in time, confirming the conclusion the vortex's role of a local sink for magnetic fields. Except for the $B_y$ component, there are no drastic changes in the orientation of the magnetic fields components at the vortices' boundary over time. 

\begin{figure*}[htp]
 \centering
        \includegraphics[width=0.9\textwidth]{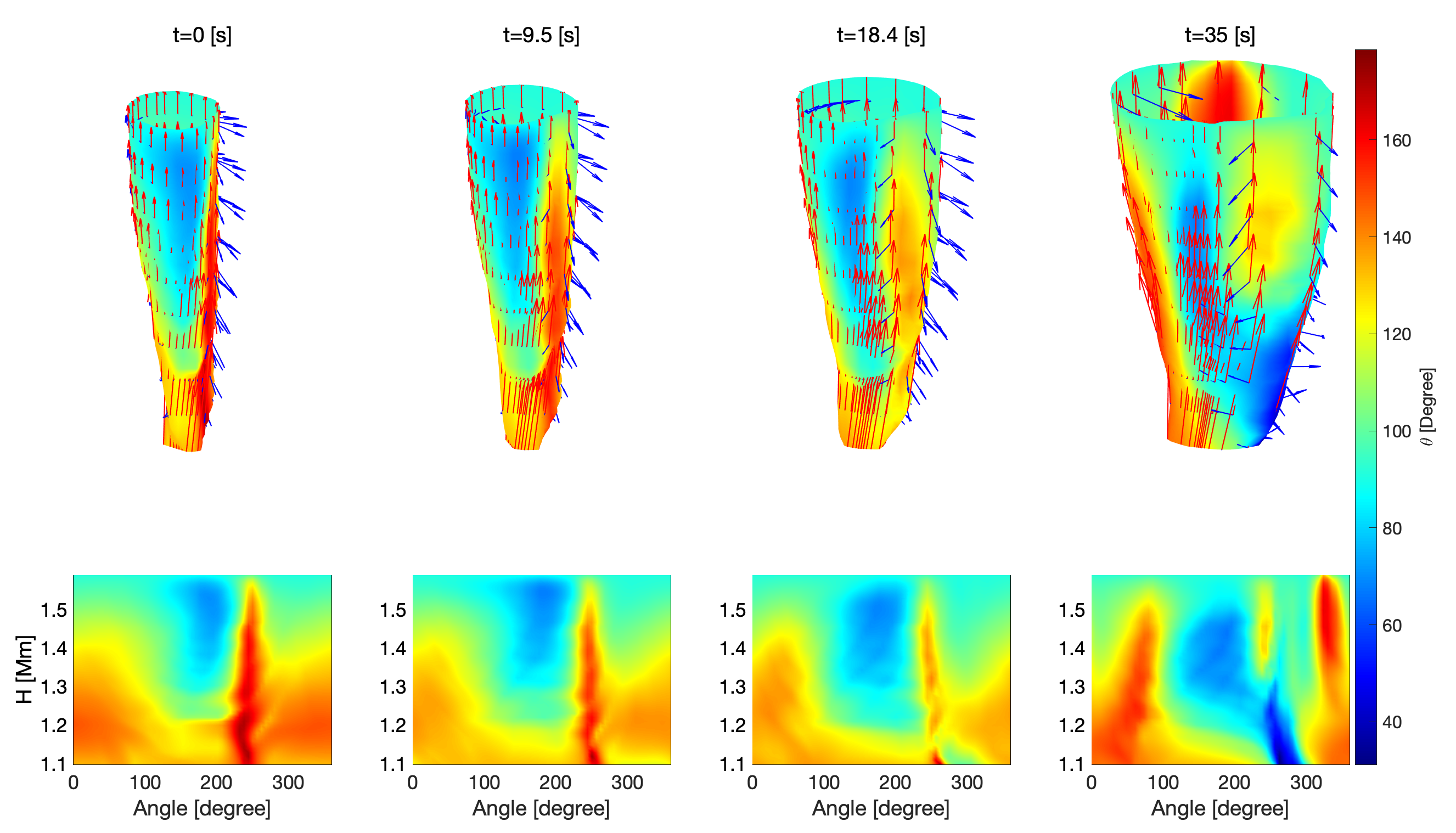}
\caption{From the left to right the four snapshots of the $R_1$ vortex are shown (the magnetic and velocity fields indicated by the red and blue arrows, respectively). The vortex surface is coloured by the angle between the velocity field and magnetic field. The same angle is shown in the bottom row in 2D. The corresponding video can be found at  \href{https://sites.google.com/sheffield.ac.uk/pdg/visualisations\#h.73wyu5789fni}{PDG visualisations} web-page.}
\label{angleR1}
\end{figure*}

\begin{figure*}[htp]
      \centering
      \includegraphics[width=0.9\textwidth]{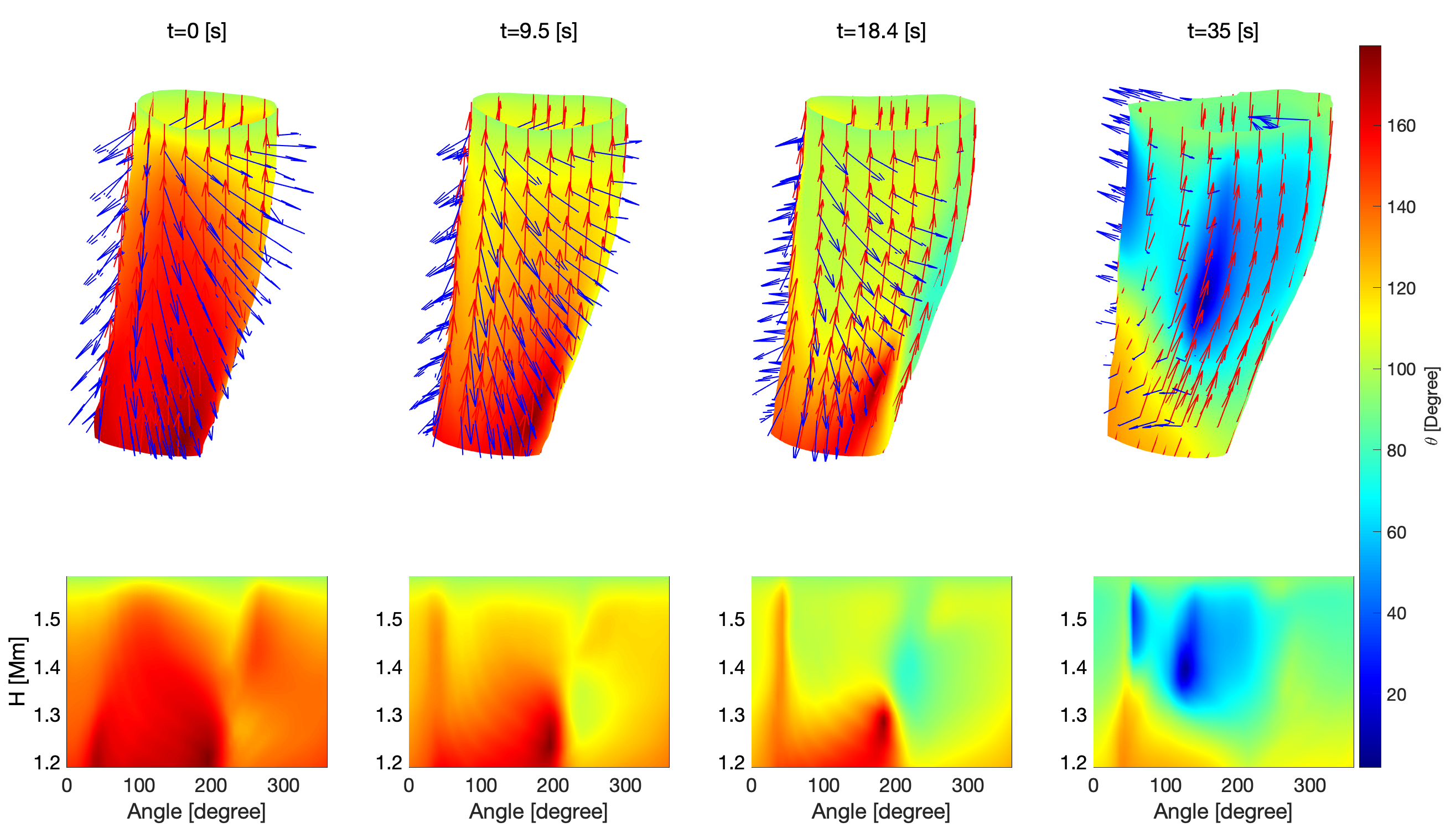} 
\caption{The same as Fig. \ref{angleR1}, but here  we present the results for vortex $L_1$. The corresponding video can be found at  \href{https://sites.google.com/sheffield.ac.uk/pdg/visualisations\#h.73wyu5789fni}{PDG visualisations} web-page.}
\label{angleL1}
\end{figure*}

One of the most interesting aspects that helps to determine the nature of the dynamics on the surface of vortices is the relative inclination of the magnetic and flow fields (see Figs.\ \ref{angleR1} and \ref{angleL1}). The magnetic field orientation is indicated by red arrows, while the direction of the flow field is given by blue arrows. The four snapshots are taken at regular time intervals, e.g. 0, 9.5, 18.4 and 35 s. In these figures the top row shows the evolution of the vortex's boundary and the directions of the two vectors fields, the bottom row shows the angle of the two vector fields measured on the surface of the vortex, where the angle is shown on the colour bar, with 180$^{\circ}$ denoting an anti-parallel orientation.
Initially, most of the lower part of the vortices presents magnetic and velocity fields directions practically anti-parallel. Looking at the arrows' orientation, it is obvious that this tends to happen in parts of the surface with strong downflow. Over time, the relative orientation of the fields tends to change and the magnetic field is mainly aligned with the plasma flow (in regions with strong upflows). The narrow and tall regions in the two snapshots of Fig.\ \ref{angleR1} correspond to a localised region where the two vector fields are anti-parallel and show an oscillatory pattern (with a characteristic length of about 100 km) that could be an indication of a wave propagation in an opposite direction to the magnetic field, advected downward by the flow (see also the fourth column of Fig.\ \ref{fig:velocity}). 

The oscillatory pattern visible in Fig.\ \ref{angleR1} is very localised. Due to the present resolution of the numerical domain, the full analysis of the nature of these wave patterns can not be performed. We speculate that these waves might be slow magnetoacoustic waves, but further high-resolution simulations and methodology improvement are needed to clarify this point. In terms of methodology, it will require incorporation of the coordinate system with the direction along the magnetic field, and perpendicular to the vortex surface to be able to project the vector field parameters. This approach will be useful for the identification of velocity and magnetic field perturbations along and perpendicular directions to the background magnetic field and, therefore, wave identification. Furthermore, for an accurate determination we would also need information about the internal part of the vortex, an aspect that is not addressed here.

In the case of vortex $L_1$, at the initial time, the two vector fields are anti-parallel, however, this decreases with time and the two vectors tend to be more aligned.
\begin{figure*}[htp]
 \centering
      \includegraphics[width=0.9\textwidth]{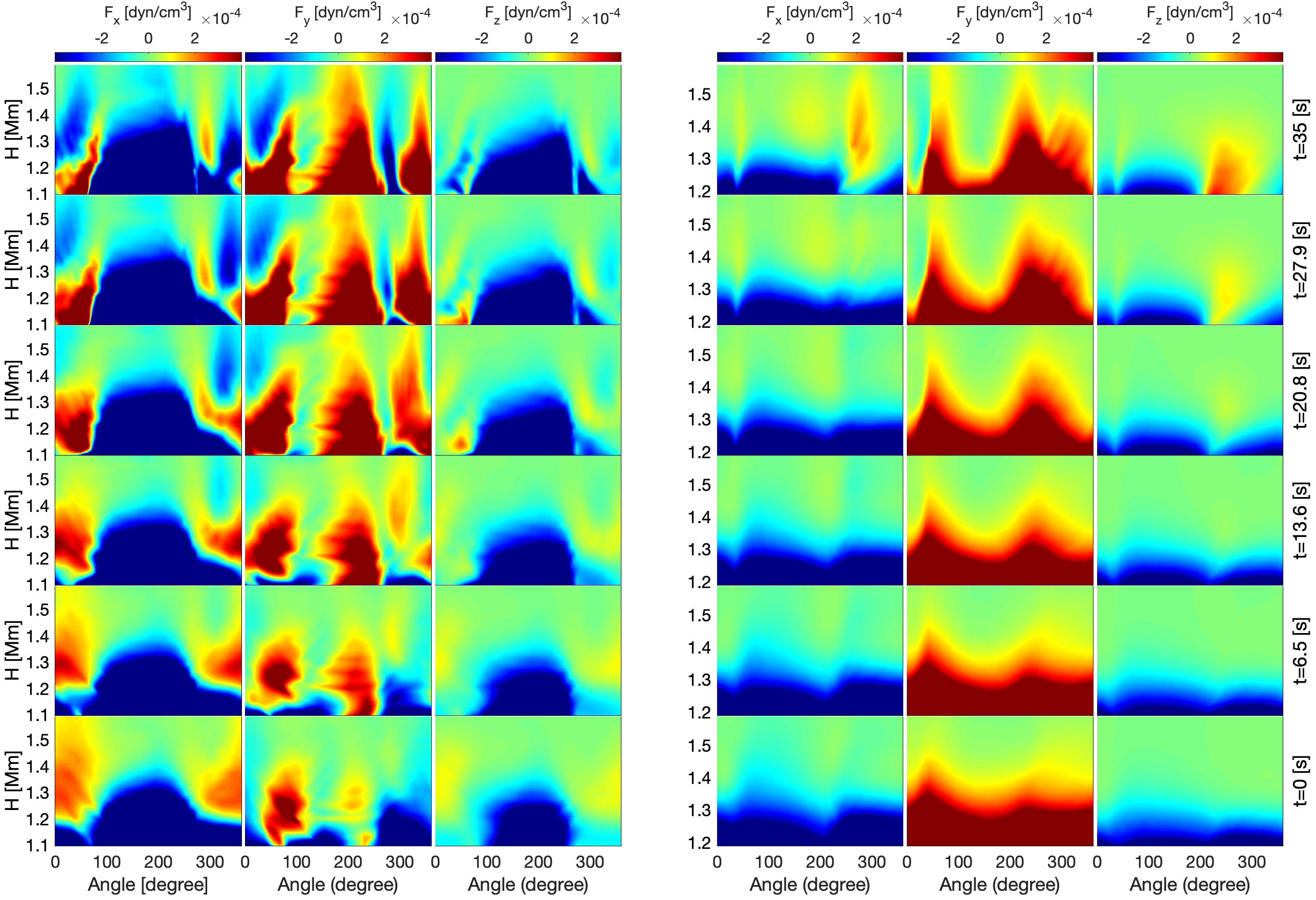}
\caption{The temporal evolution of the three components of the Lorentz force ($F_x$, $F_y$, and $F_z$) in the case of vortices $R_1$ (left panel) and $L_1$ (right panel).}
\label{fig:Lorentz}
\end{figure*}
To analyse the variation of the energy and momentum transport on the surface of the two vortices, the spatial and temporal evolution of the Lorentz force, pressure gradient and Poynting flux have been analysed. Figures  \ref{fig:Lorentz} displays (from left to right) the components of the Lorentz force ($F_x$, $F_y$, $F_z$), for the vortices $R_1$ and $L_1$, respectively. This is the magnetic restoring force oriented in the perpendicular direction to the magnetic field that acts upon changes in the magnetic field (per unit volume on the fluid) and it is defined as
\begin{equation}
{\bf F}=\frac{1}{4\pi}(\nabla\times {\bf B})\times {\bf B}.
\end{equation}
The Lorentz force introduces a magnetic pressure and also a tension along the magnetic field lines. The left panel of Fig.\ \ref{fig:Lorentz} describes the spatial and temporal evolution of the Lorentz force on the surface of the vortex $R_1$, while the right panel shows the same quantity, but in the case of vortex $L_1$. In general, the three components of Lorentz force in the two vortices demonstrate similar behaviour, with the value of the Lorentz force at the top of the analysed structures being three order of magnitude (on average) smaller than at the bottom.
The variation of the Lorentz force shows large regions where the values of its components is very small, meaning that in this region the vortex dynamics is mainly driven by hydrodynamic forces. As expected, the most dynamically changing component is the one that is in the direction of azimuthal rotation ($F_y$) and shows a significant increase during the lifetime of the vortex. The three components also indicate that for each vortex the components of the force change their direction from the lower part of the upper part of the structure. The same oscillatory pattern can also be observed in the $y$ component of the Lorentz force in the case of vortex $R_1$. In order to determine the dominant force driving the dynamics of the plasma on the surface of the two vortices, we calculate the components of the ratio of the pressure gradient force and Lorentz force as 
\begin{equation}
 C_x=-\frac{\partial p }{\partial x}/F_x,\enspace C_y=-\frac{\partial p }{\partial y}/F_y,\enspace C_z=-\frac{\partial p }{\partial z}/F_z
 \end{equation}
 
\begin{figure*}[htp]
 \centering
     \includegraphics[width=0.9\textwidth]{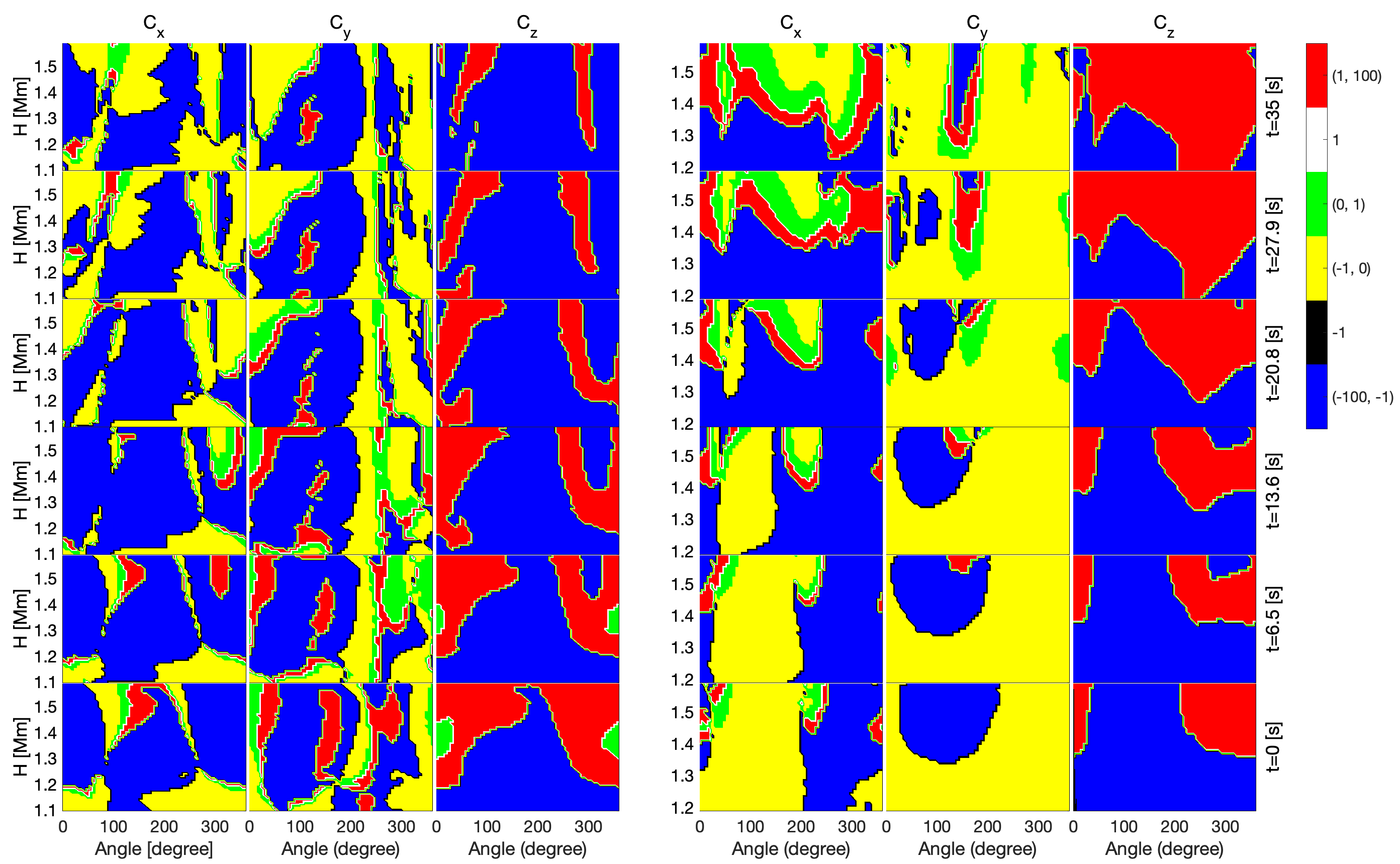}
\caption{The temporal evolution of the three components of the ratio ($C_x$, $C_y$, and $C_z$) in the case of vortices $R_1$ (left panel) and $L_1$ (right panel).}
\label{Fig:ratio}
\end{figure*}

These ratios are displayed in the Fig. \ref{Fig:ratio} for $R_1$ and $L_1$ vortices. The chosen colours denote physically distinct regions. Accordingly, the red colour denotes regions where the pressure forces dominate, but the two forces are oriented in the same direction, the blue colour denotes regions where the two forces are anti-parallel but the pressure force is higher than the Lorentz force. The green and yellow colours mark the regions where the Lorentz force is larger than the pressure force, but the two colours are denoting the cases when the two vectors are parallel and anti-parallel, respectively. Finally, the white and black colours denote the cases when the two forces are equal, but the their orientation is parallel and anti-parallel. 
From Fig.\ \ref{Fig:ratio} it follows that on the boundary surface of the vortex the two forces are never in balance. In the case of vortex $R_1$ the dynamics of the plasma in the horizontal directions show a rather complex pattern, with localised regions where the role of the dominant force is changing. When compared to the results showed for the distribution of velocity (see Fig.\ \ref{fig:velocity}), it is clear that in the case of vortex $R_1$ the plasma motion is mainly driven by pressure forces and the two forces are pointing, in general, in different directions. In time, the role of pressure forces diminish. In the vertical direction, there is also the tendency of pressure gradient forces being larger than the Lorentz force. The dynamics on the surface of vortex $L_1$ shows similar complexity thanks to the interplay of the two forces. Comparing these results with the findings shown in Fig.\ \ref{fig:velocity} (left panel), it is clear that initially the flows in the positive $x$ direction are driven by pressure forces, while the flows in the negative direction are driven by magnetic forces. Looking at the second panel of Fig. {\ref{Fig:ratio}} we can conclude that in the $y$ direction the two forces are almost always anti-parallel and motion in the positive/negative directions are driven by magnetic/pressure forces. Finally, the motion in the vertical direction shows an interesting feature, when the alignment of the two forces is changing in time from being mainly anti-parallel, to parallel, but pressure forces dominate everywhere.

\begin{figure*}[htp]
 \centering
     \includegraphics[width=0.9\textwidth]{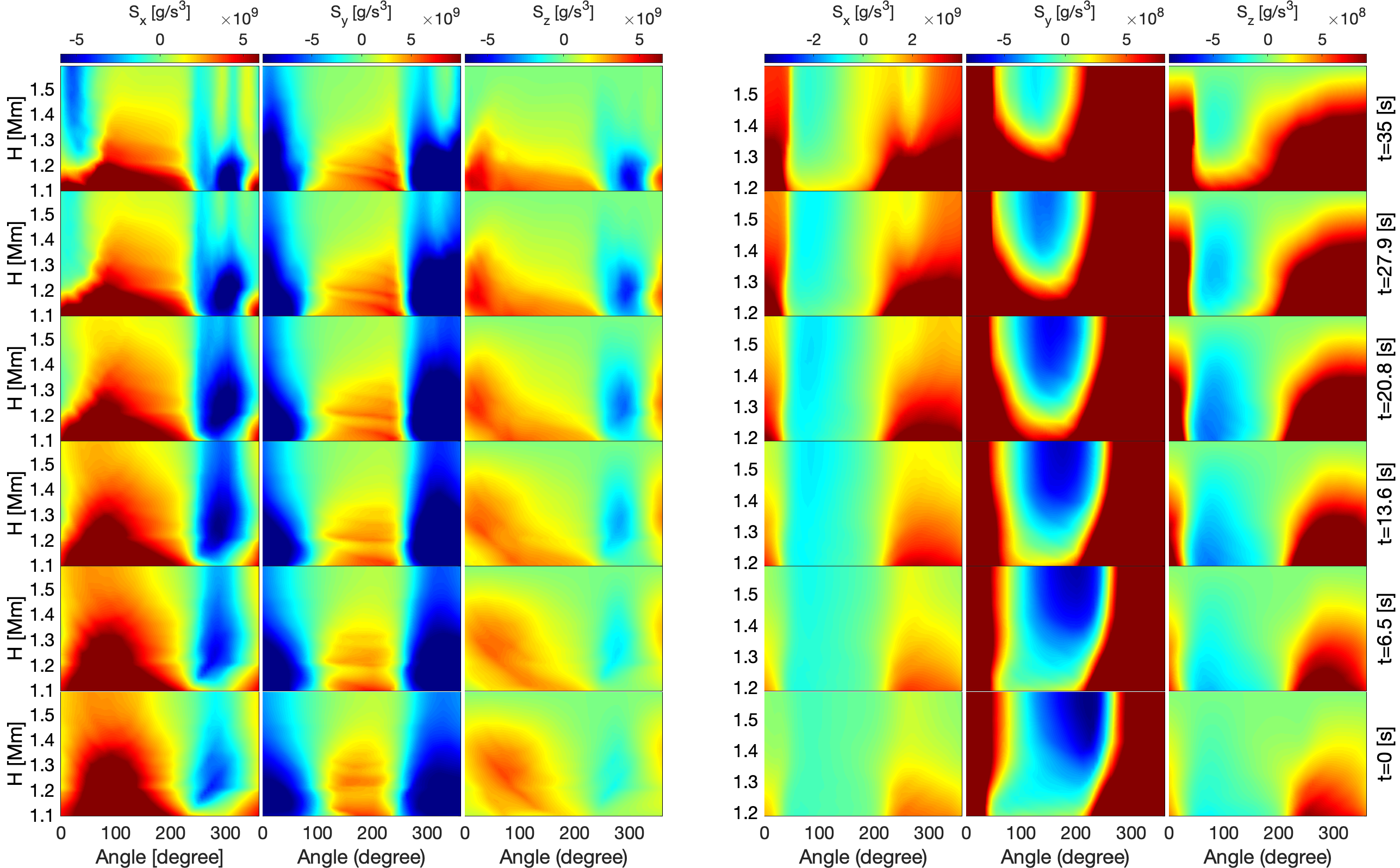} 
\caption{The temporal evolution of the three components of Poynting flux ($S_x$, $S_y$, and $S_z$) in the case of vortices $R_1$ (left panel) and $L_1$ (right panel).}
\label{Fig:flux}
\end{figure*}

The MHD approximation of the Poynting flux is defined as the energy transferred by a wave across a unit area at any instant time and it is given by
\begin{equation}
{\bf S}=\frac{1}{4\pi}{\bf B}\times ({
\bf v}\times {\bf B})=\frac{1}{4 \pi} \left[ \mathbf{v}(\mathbf{B} \cdot \mathbf{B}) - \mathbf{B}(\mathbf{B} \cdot \mathbf{v})\right]. 
\end{equation}
The first term in the above equation is oriented in the direction of the velocity field, and it is proportional to the square of the magnetic field intensity. In contrast, the second term is in the same direction as the magnetic field, and it is proportional to the alignment of the magnetic and velocity fields. 
The evolution of the Poynting flux components ($S_x$, $S_y$ and $S_z$) for the vortex $R_1$ is shown in the left panel of Fig. \ref{Fig:flux} and the horizontal components show high correlation with the pattern of the plasma flow velocity shown in Fig. \ref{fig:velocity}. The fact that these components have positive and negative values is attributed to the directions of the plasma flow, but in any case the direction of the energy flow agrees with the direction of the flow. It is interesting to note that the vertical component of the Poynting flux shows large regions where this quantity is close to zero, meaning that the electromagnetic energy is distributed mainly in the horizontal direction confirming the findings of  \cite{Silva_2022}. 
In the case of vortex $L_1$ (see Fig. \ref{Fig:flux} right panel) the magnitude and the direction of the Poynting flux agrees with the direction of the plasma flow. Again, initially the Poynting flux is mainly horizontal, but in time, the vertical component increases thanks to the increase in the plasma flow seen in Fig. \ref{fig:velocity}. It is  interesting to note the same periodic variation of the $y$ component of the Poynting flux similar to the variation of the $y$ component of the Lorentz force, and the characteristic length of these variation are similar. 

\section{Discussions and conclusions}

The current study was dedicated to the analysis of physical parameters on the surface of vortices in magnetoconvection numerical data which correspond to the dynamics in the lower part of the solar atmosphere. The comparison of radiative magnetoconvection simulations with the real observations has been reported in the number of earlier studies, e.g. \cite{2003ApJ...597L.173S,2004A&A...427..335S,2005AmMin..90.1913K,2007A&A...469..731S} to name but a few. Notably, \cite{2012A&A...539A.121B} have compared results from a number of photospheric magneto-convection simulations produced by different codes and found that the results reliably represent the finest details of the observed solar radiation, including its temporal and spatial scales. Here, the three-dimensional vortex structure has been determined using Lagrangian analysis and the vortex boundaries were defined by means of isosurfaces of LAVD contours as previously suggested by \cite{Haller_a1}. Physically, the use of this technique means that all the particles on the surface boundary undergo the same intrinsic dynamic rotation. In other words, in the time interval used to compute the LAVD field, the fluid elements at the vortex boundary experience the same bulk rotation relative to the mean rigid body rotation of the fluid. The radius of the detected Lagrangian vortices increases with height, as also found for instantaneous vortex detection \citep{Silva_2020}. For the selected vortices, we see the surfaces change in size while keeping their topology. In this paper, the LAVD approach was applied to a small subset of the numerical domain. Due to the fact that the procedure is automated, it may be used in large regions, too. This may be accomplished rather simply by selecting a constant convexity deficiency value and constantly searching for the LAVD isosurface that begins at the same height.

The variation of physical parameters on the surface of the examined vortices show a very complex behaviour. The LAVD field was used to define vortex centres and their boundaries based on the fluid elements advection. The case of vortices presented in the current study, global changes are not occurring for all physical parameters. It is clear that the variation of density, pressure, particular magnetic field components are, to a very large extent, changing globally (see Figs. \ref{Fig:statevariables} and \ref{fig:magnetic}), however, the general characteristic of investigated vortices is that physical parameters describing the state of the plasma and quantities describing dynamics and energetics have a much more local character, meaning that changes occur locally (see, e.g. the variation of temperature, plasma-$\beta$, velocity and Lorentz force). The analysis of the Poynting flux components (see Fig. \ref{Fig:flux}) reveals that the energy flows are mostly present in the horizontal direction as previously discussed by \cite{Silva_2022}.

Due to the simultaneous presence of the oppositely oriented flows, the solar atmospheric vortices are mixing plasma predominantly in the vertical direction. The forces that drive the plasma dynamics on the vortices surfaces show a high degree of inhomogeneity (see Fig. \ref{Fig:ratio}, for example). We anticipate that the increase of vertical velocity component ($v_z$) is related to the net force balance in the vertical direction. The pressure gradient is always pointing upwards and the gravity is always pointing downwards. Therefore, although the pressure gradient tends to dominate over the $z$-component of Lorentz force, from results presented in Fig. \ref{fig:Lorentz} it follows that the changes in the vertical velocity are actually driven by the Lorentz force. Indeed, for both $R_1$ and $L_1$, we have deceleration and acceleration of vertical flow in regions where the Lorentz force is increasing downwards and upwards respectively. The pressure gradient dominance is expected since we are studying the vortex boundary, i.e. our analysis concerns the outermost region where the flow dynamics is more effectively affected by the vortical motions. 

The structures we studied also present dynamics whose driver undergoes a transition, from being driven by kinetic forces, to a motion that is driven by magnetic forces. Therefore, the magnetic nature of the vortex, observed in previous studies \citep{Shelyag_2012,Kitiashvili_2013,Silva_2020}, is not prevailing over the kinematic dynamics.
Our results suggest the Lagrangian vortices recovered from the magneto-convection simulations do not rotate as a rigid body. This particular rotation will lead to the different angular velocities at different heights of the vortex. The analysis of the angular velocity profiles as a function of radius for the vortices identified in the magnetoconvection numerical simulations were presented in a recent investigation by \cite{Silva_2020}.


Future works will focus on the description of plasma and field parameters inside the magneto-convection Lagrangian vortices. Recent study by \citep{Battaglia2021} has evidenced the propagation of Alfv\'en pulses inside a vortex structure. Our results suggest that we may have some signatures of wave propagation on the LAVD-identified vortex surface as periodic changes were observed for some variables (see Figs. \ref{fig:Lorentz} and \ref{Fig:flux}). Therefore, the methodology used in our paper to determine the values of physical parameters and their changes in time is suitable for the identification of the dominant physical parameters that drive the wave propagation. 

In general, it is conceivable to compare 3D vortex surfaces (reconstructed from numerical simulations) with their counterparts from high-resolution observations of the solar atmosphere, but this is not straightforward. The main difficulty is related to the reconstruction of the horizontal velocity field from photospheric intensity observations. The available local correlation tracking \citep[LCT;][]{November1988}, FLCT \citep{Fisher2008} and recently developed DeepVel \citep{Asensio_Ramos2017} and Multi-Scale Deep Learning \citep{Ishikawa2021} methodologies have a number of limitations which may lead to the not fully accurate interpretation of the real physical plasma flows (see the above references for more details). The second issue is related to the data sets available for the analysis. The precise vortex spatial structure can be obtained from the magneto-convection simulations, but in observations only few horizontal slices (which correspond to the integral signatures of the spectral lines formed at different heights) can be used for analysis. Therefore, 3D reconstructions of the vortex dynamics which are based on observational data sets may imply errors that would influence the obtained results. However, it is worth mentioning that even in this case it is possible to directly compare the temporal and spatial evolution of observable and numerically simulated vortex parameters, e.g. radial dependence of velocity or magnetic field strength by taking 2D horizontal slices. Our research on this aspect is currently in progress.


\keywords{editorials, notices --- 
miscellaneous --- catalogs --- surveys}


\acknowledgments
YA acknowledges the Deanship of Scientific Research
(DSR), Umm Al-Qura University (Saudi Arabia), for the financial
support. VF, GV, IB, and SSAS are grateful to The Royal Society, International Exchanges Scheme, collaboration with Brazil
(IES\textbackslash R1\textbackslash 191114). VF and GV are grateful to Science and Technology Facilities Council (STFC) grant ST/V000977/1
and to The Royal Society, International Exchanges Scheme, collaboration with Chile (IE170301). VF would like to
thank the International Space Science Institute (ISSI) in Bern, Switzerland, for the hospitality provided to the members of the team on ‘The Nature and Physics of Vortex Flows in Solar Plasmas’. This research has also received financial support from the European Union’s Horizon 2020 research and innovation program under grant agreement No. 824135 (SOLARNET). This research have been undertaken with the assistance of resources and services from the National Computational Infrastructure (NCI), which is supported by the Australian Government. Some of the results were obtained using the OzSTAR national facility at Swinburne University of Technology. The OzSTAR program receives funding in part from the Astronomy National Collaborative Research Infrastructure Strategy (NCRIS) allocation provided by the Australian Government. This research was supported partially by the Australian Government through the Australian Research Council's Discovery Projects funding scheme (project DP160100746) and through Future Fellowship FT120100057 awarded to Dr S. Shelyag.

\bibliography{Aljohani_etal_vortex1}{}
\bibliographystyle{aasjournal}




\end{document}